\documentclass[fleqn,usenatbib]{mnras}

\usepackage{newtxtext,newtxmath}

\usepackage[T1]{fontenc}
\usepackage{ae,aecompl}

\usepackage{graphicx}	
\usepackage{amsmath}	
\usepackage{amssymb}	
\usepackage{multirow}

\newcommand{\fb}{MKT~J170456.2$-$482100} 
\newcommand{\kstar}{TYC~8332-2529-1}
\newcommand{\gx}{GX339$-$4}
\newcommand{\trap}{\textsc{TraP}}
\newcommand{\psr}{PSR\,J1703$-$4851}

\def\arcsec{\hbox{$^{\prime\prime}$}}

\emergencystretch=\maxdimen
\title[\fb: the first MeerKAT transient]{\fb: the first transient discovered by MeerKAT}

\author[L. N. Driessen et al.]{L. N. Driessen,$^{1}$\thanks{E-mail: Laura@Driessen.net.au (LND)}
I. McDonald,$^{1}$
D. A. H. Buckley,$^{2}$
M. Caleb,$^{1}$
E. J. Kotze,$^{2,20}$
\newauthor
S. B. Potter,$^{2}$
K. M. Rajwade,$^{1}$
A. Rowlinson,$^{3,4}$
B. W. Stappers,$^{1}$
E. Tremou,$^{5}$
\newauthor 
P. A. Woudt,$^{6}$
R. P. Fender,$^{6,7}$
R. Armstrong,$^{6,8}$
P. Groot,$^{2,6,9}$
I. Heywood,$^{7,10}$
\newauthor 
A. Horesh,$^{11}$
A. J. van der Horst,$^{12,13}$
E. Koerding,$^{9}$
V. A. McBride,$^{2,14,15}$
\newauthor 
J. C. A. Miller-Jones,$^{16}$
K. P. Mooley$^{17,18,19}$
and R. A. M. J. Wijers$^{3}$
\\ \\
$^{1}$Jodrell Bank Centre for Astrophysics, Department of Physics and Astronomy, The University of Manchester, Manchester, M13 9PL, UK\\
$^{2}$South African Astronomical Observatory, PO Box 9, Observatory 7935, South Africa\\
$^{3}$Anton Pannekoek Institute, University of Amsterdam, Postbus 94249, 1090 GE, Amsterdam, The Netherlands\\
$^{4}$Netherlands Institute for Radio Astronomy (ASTRON), Oude Hoogeveensedijk 4, 7991 PD, Dwingeloo, The Netherlands\\
$^{5}$AIM, CEA, CNRS, Universit\'e Paris Diderot, Sorbonne Paris Cit\'e, Universit\'e Paris-Saclay, F-91191 Gif-sur-Yvette, France\\
$^{6}$Inter-University Institute for Data Intensive Astronomy, Department of Astronomy, University of Cape Town, Private Bag X3,\\Rondebosch 7701, South Africa\\
$^{7}$Department of Physics, Astrophysics, University of Oxford, Denys Wilkinson Building, Keble Road, Oxford OX1 3RH, UK\\
$^{8}$South African Radio Astronomy Observatory, 2 Fir Street, Black River Park, Observatory, Cape Town 7925, South Africa\\
$^{9}$Department of Astrophysics/IMAPP, Radboud University Nijmegen, P.O. Box 9010, 6500 GL Nijmegen, The Netherlands\\
$^{10}$Department of Physics and Electronics, Rhodes University, PO Box 94, Grahamstown 6140, South Africa\\
$^{11}$Racah Institute of Physics, The Hebrew University of Jerusalem, Jerusalem 91904, Israel\\
$^{12}$Department of Physics, The George Washington University, 725 21st Street NW, Washington, DC 20052, USA\\
$^{13}$Astronomy, Physics and Statistics Institute of Sciences (APSIS), 725 21st Street NW, Washington, DC 20052, USA\\
$^{14}$Department of Astronomy, University of Cape Town, Private Bag X3, Rondebosch 7701, South Africa\\
$^{15}$IAU Office of Astronomy for Development, Cape Town, 7935, South Africa\\
$^{16}$International Centre for Radio Astronomy Research -- Curtin University, GPO Box U1987, Perth, WA 6845, Australia\\
$^{17}$Department of Physics, University of Oxford, Keble Road, Oxford OX1 3RH, UK\\
$^{18}$National Radio Astronomy Observatory, Socorro, NM 87801, USA\\
$^{19}$Caltech, 1200 E. California Blvd. MC 249-17, Pasadena, CA 91125, USA\\
$^{20}$Southern African Large Telescope, P.O.Box 9, Observatory, 7935, South Africa\\
}
\date{Accepted 2019 October 18. Received 2019 October 18; in original form 2019 August 14}

\pubyear{2019}

\begin{document}
\label{firstpage}
\pagerange{\pageref{firstpage}--\pageref{lastpage}}
\maketitle

\begin{abstract}
We report the discovery of the first transient with MeerKAT, \fb, discovered in ThunderKAT images of the low mass X-ray binary GX339$-$4.
\fb\, is variable in the radio, reaching a maximum flux density of $0.71\pm0.11\,\mathrm{mJy}$ on 2019 Oct 12, and is undetected in 15 out of 48 ThunderKAT epochs.
\fb\, is coincident with the chromospherically active K-type sub-giant \kstar, and $\sim18\,\mathrm{years}$ of archival optical photometry of the star shows that it varies with a period of $21.25\pm0.04\,\mathrm{days}$.
The shape and phase of the optical light curve changes over time, and we detect both X-ray and UV emission at the position of \fb, which may indicate that \kstar\, has large star spots. Spectroscopic analysis shows that \kstar\, is in a binary, and has a line-of-sight radial velocity amplitude of $43\,\mathrm{km\,s^{-1}}$.
We also observe a spectral feature in anti-phase with the K-type sub-giant, with a line-of-sight radial velocity amplitude of $\sim12\pm10\,\mathrm{km\,s^{-1}}$, whose origins cannot currently be explained.
Further observations and investigation are required to determine the nature of the \fb\, system.
\end{abstract}



\section{Introduction}

The radio sky contains many variable and transient sources, often found 
in follow-up observations of transients detected at
at other wavelengths such as optical, gamma-ray and X-ray \citep[e.g.][]{1994IAUC.6006....1S,2011Natur.476..425Z,2012ApJ...746..156C,2013ApJ...778...63H,2015ApJ...815..102F,2016Natur.537..374M,2017Sci...358.1579H,2019MNRAS.486.2721B}. Blind searches for radio transients using interferometers present many challenges, particularly modest field of view (FoV) and limited observing cadence \citep[e.g.][]{2013PASA...30....6M,2016ApJ...818..105M,2018ApJ...857..143M}.
With current wide FoV ($\gtrsim1\,\mathrm{square\,degree}$) instruments such as MeerKAT \citep{2018ApJ...856..180C}, the Australian Square Kilometer Array Pathfinder \citep[ASKAP;][]{2008ExA....22..151J,2012SPIE.8444E..2AS}, APERTIF \citep[][]{2017arXiv170906104M}, the LOw Frequency Array  \citep[LOFAR;][]{2013A&A...556A...2V}, and the Murchison Wide Field Array \citep[MWA;][]{2012rsri.confE..36T}, surveying large areas of sky with various cadences and improved sensitivity is now possible. These new instruments could result in the discovery of tens to hundreds of transients \citep[e.g.][]{2015aska.confE..62O}.

Radio transients are commonly divided into two categories: coherent and incoherent \citep[e.g.][]{2015MNRAS.446.3687P}; and both types of transient are investigated in the time domain with high-time resolution (milliseconds or less), and in image plane observations with a range of integration timescales.
In this publication we will focus on image plane searches. Current image plane transient searches include the Amsterdam-ASTRON Radio Transients Facility and Analysis Centre \citep[AARTFAAC;][]{2016JAI.....541008P,2019MNRAS.482.2502K}, and the ASKAP Survey for Variables and Slow Transients \citep[VAST;][]{2013PASA...30....6M}. Large surveys such as the Very Large Array (VLA) Sky Survey \citep[VLASS;][]{2019arXiv190701981L} are also being used to search for transients \citep{hallinan2013transient}.
It was originally theorised that image plane, low-frequency transient searches would detect many transient radio sources, but to date only one transient each has been found with LOFAR \citep{2016MNRAS.459.3161C,2016MNRAS.456.2321S}, the Long Wavelength Array \citep[LWA,][]{2019ApJ...874..151V} and the MWA \citep{2017MNRAS.466.1944M}, and no transients have been found with the VLA Low Band Ionospheric and Transient Experiment \citep[VLITE;][]{2016ApJ...832...60P}. The rate of low-frequency Galactic
transients may be higher, as inferred from the Galactic Center Radio
Transients detected by VLA and Giant Metrewave Radio Telescope \citep[GMRT, e.g.][]{2005Natur.434...50H,2009ApJ...696..280H,2010ApJ...712L...5R}. At higher frequencies, a VLA search for transients at 5 GHz only found a single transient candidate \citep{2011ApJ...740...65O}, while the Caltech-NRAO Stripe 82 Survey Pilot \citep[CNSS;][]{2016ApJ...818..105M} at $3\,\mathrm{GHz}$ detected two transients, and several transients have been found in the full CNSS survey (Mooley et al. 2019, in preparation). \citet{2007ApJ...666..346B} searched 944 epochs over 22 years of VLA 5 and 8 GHz observations and found 10 new transients; however, more than half of these were found to be artefacts by \citet{2012ApJ...747...70F}. More recently, 9 potential variable sources were detected using ASKAP \citep{2018MNRAS.478.1784B}. This means that only a few radio transients have been discovered in blind image plane transient surveys, despite expectations for many new discoveries. These results\footnote{A comprehensive list of blind radio transient
searches can be found at
\href{http://www.tauceti.caltech.edu/kunal/radio-transient-surveys/index.html}{http://www.tauceti.caltech.edu/kunal/radio-transient-surveys/index.html}.}
have highlighted the importance of wide-field, sensitive searches, at suitable frequencies, for maximising the
yield of radio transients \citep[see also][]{2011ApJ...742...49T,2011MNRAS.412..634B}.

One type of radio transient expected to be found in image plane transient searches is flares from stars and stellar systems \citep[see e.g.][for a summary]{2008arXiv0801.2573O}.
Radio flare stars are usually M-type dwarf stars that emit coherent radio bursts on times scales of minutes to hours. Recently, \citet{2019ApJ...871..214V} detected 22 coherent radio bursts from M dwarfs using the VLA at 300 MHz and 1-6 GHz, and \citet{2019arXiv190606570Z} detected several pulses from the M-dwarf UV Ceti with ASKAP. As well as flare stars, binary systems such as RS Canum Venaticorum (RS CVn), Cataclysmic Variables (CVs) and symbiotic binaries are known to flare in the radio. RS CVn are binary systems consisting of a late-type giant or sub-giant star with a late-type main sequence star companion \citep[e.g.][]{1976ASSL...60..287H,1997ApJS..113..131C,2003A&A...403..613G}. RS CVn are chromospherically active and the giant or sub-giant rotates quasi-synchoronously with the orbital period. Periods for RS CVn are typically 1-20 days, and radio flares on RS CVn systems have been observed to last up to a few days \citep[e.g.][]{1987A&A...186..241W,1993MNRAS.260..903T}. RS CVn emit in both the radio and X-ray while in quiescence \citep[e.g.][]{1996IrAJ...23..137G}. CVs are binary systems with a white dwarf primary accreting matter from Roche lobe overflow of the secondary star, usually a main-sequence star \citep[e.g.][]{2016MNRAS.463.2229C}. Dwarf novae from magnetic CVs have been observed to exhibit radio outbursts that can last for weeks, with rapid radio flaring on timescales less than an hour \citep{2017MNRAS.467L..31M}. Symbiotic binaries also have a white dwarf primary, but the companion is a red giant star and the orbit is wide. In these systems, mass is accreted on to the white dwarf via stellar winds \citep{2015aska.confE..62O}. Radio variability over timescales of a few hours have been detected on symbiotic binaries, for example RX Puppis \citep{1977ApJ...211..547S}. AR Scorpii is another stellar binary that is observed in the radio, it consists of a white dwarf primary and an M-dwarf companion. AR Scorpii has an orbital period of 3.56 hours, is observed to pulsate in the optical, radio and X-ray with a period of 1.97 minutes \citep{2016Natur.537..374M,2018ApJ...853..106T}, and is highly polarised \citep[e.g.][]{2017NatAs...1E..29B}. Radio flaring stellar systems vary on a variety of time scales, making it difficult to detect when these systems are in outburst \citep[e.g.][]{2008ApJ...674.1078O}. This means that wide-field monitoring is an important method for discovering and investigating these sources in the radio.

MeerKAT is the (more) Karoo Array Telescope \citep{2018ApJ...856..180C} consisting of 64, 13.96\,m dishes in South Africa. MeerKAT has a FoV of over a square degree at 1.4\,GHz, which makes it an excellent instrument for searching for radio transients, such as stellar flares. ThunderKAT is the MeerKAT Large Survey Project (LSP) investigating and searching for transients in the image plane \citep{2017arXiv171104132F}. ThunderKAT is directly observing transient sources such as X-ray binaries, CVs, and gamma-ray bursts, and is commensally searching for radio transients. ThunderKAT has committed to observing the low mass X-ray binary \gx\, every week for five years. With a FoV of over a square degree, weekly observations, and hundreds of sources in the field, this is an excellent opportunity to search for transient and variable radio sources that vary on many different timescales. 
In this paper we will present the serendipitous discovery of \fb, the first radio transient discovered commensally with MeerKAT, in the \gx\, field. In Section\,\ref{sec: meerkat obs} we will present the radio observations of the source and the method of detection. In Section\,\ref{sec: optical ID} we will identify the optical counterpart to \fb. In Sections\,\ref{sec: optical photometry}, \ref{sec: optical spectroscopy}, \ref{sec: pulsation searches}, \ref{sec: UV obs} and \ref{sec: X-ray obs} we will present the optical photometry, optical spectroscopy, radio pulsation searches, UV photometry, and X-ray photometry of the source respectively. In Sections\,\ref{sec: discussion} and \ref{sec: conclusion} we will discuss our findings and conclude.

\section{MeerKAT radio observations}
\label{sec: meerkat obs}

ThunderKAT \citep{2017arXiv171104132F} first observed the \gx\, (Tremou et al. in prep) field with 16 dishes on 2017 November 11 during commissioning.
It then observed the field with all 64 dishes for the first time on 2018 April 14, and it began weekly monitoring in September 2018. 
We present data from 48 epochs, 46 epochs from weekly monitoring plus the 2017 November 11 and 2018 April 14 epochs.
The \gx\, field is observed using the L-band ($900$-$1670\,\mathrm{MHz}$) receiver in full polarisation mode, which has a bandwidth of $856\,\mathrm{MHz}$, a central frequency of $1284\,\mathrm{MHz}$, and 4096 frequency channels.
The observations are typically 10-15\,minutes in duration, with a minimum integration time of 8\,seconds.
The phase calibrator (1722$-$554) is observed for $\sim2$\,minutes before or after observing the field, and the band-pass and flux calibrator (1934$-$638) is observed for 5\,minutes at the beginning of the observing block.

Each observation is first flagged using AOFlagger\footnote{\href{https://sourceforge.net/projects/aoflagger/}{https://sourceforge.net/projects/aoflagger/}} \citep{2010offringa} and calibrated by following standard procedures (a priori phase correction, antenna delays and band-pass corrections) using the Common Astronomy Software Application\footnote{\href{https://casa.nrao.edu/}{https://casa.nrao.edu/}}  \citep[CASA;][]{2007mcmulin}. As the data volume is large, the data are binned to reduce the number of channels from 4096 to 512. The full ThunderKAT pipeline will be presented in a future publication.

The data are imaged using the new wide-band, wide-field imager, DDFacet \citep{2018tasse}. DDFacet is based on a co-planar faceting scheme, and it takes generic direction dependent effects into account. This is important as MeerKAT has a very wide ($\sim1$\, square degree) FoV. The imaging was performed by deconvolving over 4 frequency subbands using the \textsc{SSDclean} deconvolution algorithm and a Briggs weighting scheme (robust=$-$0.7). To correct for considerable artefacts from bright sources, self-calibration was performed using the KillMS\footnote{\href{https://github.com/saopicc/killMS}{https://github.com/saopicc/killMS}} software that accompanies DDFacet. The image quality was optimised by using the Complex Half-Jacobian Optimisation for N-directional EStimation \citep[\textsc{CohJones};][]{2015tasse} algorithm to correct for direction-dependent effects. The \textsc{CohJones} algorithm solves for scalar Jones matrices in a user-defined number of directions; three directions were used for the \gx\, field. For the final images of the field the median synthesised beam in L-band is 5$\arcsec$ $\times$ 4$\arcsec$. The images typically have an overall root-mean-square (RMS) noise of $\sim31.7\,\mathrm{\upmu\,Jy\,beam^{-1}}$.  We have found that there is a flux-dependent underlying systematic flux density fluctuation of up to the $10$ per cent in the light curves of sources in this field, likely caused by flux density variation in the secondary calibrator source. This does not affect the results of the analysis in this work; however, any small-scale variability visible in the MeerKAT light curves presented here should not be over-interpreted.

\subsection{Transient detection pipeline}
\label{sec:trap}

\begin{figure}
 \includegraphics[width=\columnwidth]{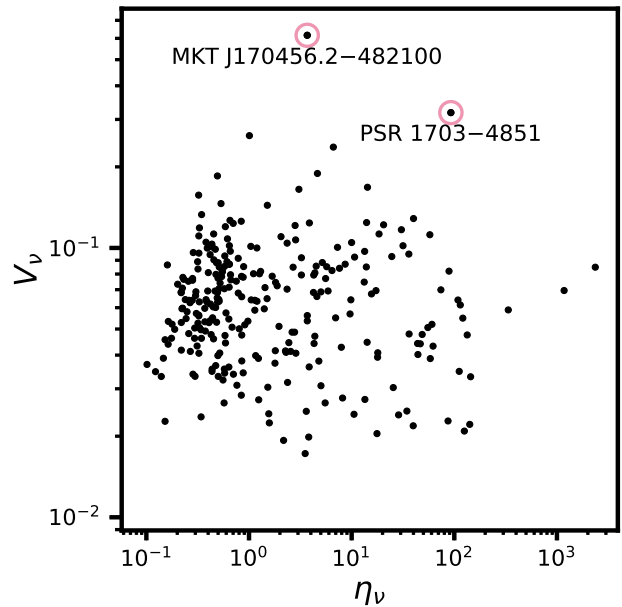}
 \caption{The variability parameters for all of the sources detected in the field of \gx. \psr\, and \fb\, (labelled) are clear outliers from the population of sources.}
 \label{fig:trap_plot}
\end{figure}

In December 2018 the images from the regular monitoring of \gx\, were used to trial the LOFAR \textsc{Transient Pipeline} \citep[\trap, Release 4.0;][]{Swinbank2015} for detecting variable and transient sources in MeerKAT observations. \trap\, automatically processes a time series of images by finding sources, determining source associations, and constructing light curves. The light curves are then used to calculate two variability parameters: $\eta$  and $V$. The $\eta$ parameter is the reduced $\chi^{2}$ value for a fit to a stable source, higher $\eta$ values show increased variation from a constant flux density while lower $\eta$ values indicate that the flux density is stable. The $V$ parameter is the co-efficient of variability or the modulation parameter, which is defined as the ratio between the standard deviation in the flux density and the mean flux density value. A higher $V$ value indicates larger fractional variation in flux density values. For further information on the processing steps and the calculation of the variability parameters refer to \cite{Swinbank2015}.

The default \trap\, settings were used to analyse the \gx\, field\footnote{For further details about the \trap\, capabilities, refer to the \trap documentation at \href{https://tkp.readthedocs.io/en/latest/}{https://tkp.readthedocs.io/en/latest/} \citep{TraP}}.
A source detection threshold of $8\sigma$ was used, and the source finder was forced to use a Gaussian shape consistent with the synthesised beam of the images to search for sources.
This was done to prevent the source finder from detecting extended emission, as variable sources are typically expected to be point sources.
The first trial use of \trap\, on the \gx\, field was highly successful, the distributions of the variability parameters $\eta$ and $V$ are shown in Figure\,\ref{fig:trap_plot}. 
An outlier in both $\eta$  and $V$ was found to coincide with the known pulsar \psr, labelled in Figure\,\ref{fig:trap_plot}, at Right Ascension and Declination of 17h03m54.53(2)s and $-$48d52m01.0(5)s  respectively \citep[J2000;][]{2019MNRAS.484.3691J}. This is a known mode-changing pulsar \citep{2007MNRAS.377.1383W}, and this mode changing causes the flux density integrated over $\sim10\,\mathrm{minutes}$ to vary from epoch to epoch. An interesting source, \fb, was also identified as an outlier to the distribution of the $V$ variability parameter. It was confirmed as a variable source by visual inspection of the light curve and images. 
At the time of discovery the variability parameters for \fb\, were $\eta=3.7$ and $V=0.62$, while the point sources near \fb\, are consistent with stable sources with $0.15\leq \eta \leq0.5$ and $0.04\leq V \leq0.09$.
We note that there are other outliers in Figure\,\ref{fig:trap_plot}, these will be investigated and presented in a future publication.

\begin{figure*}
 \includegraphics[width=\textwidth]{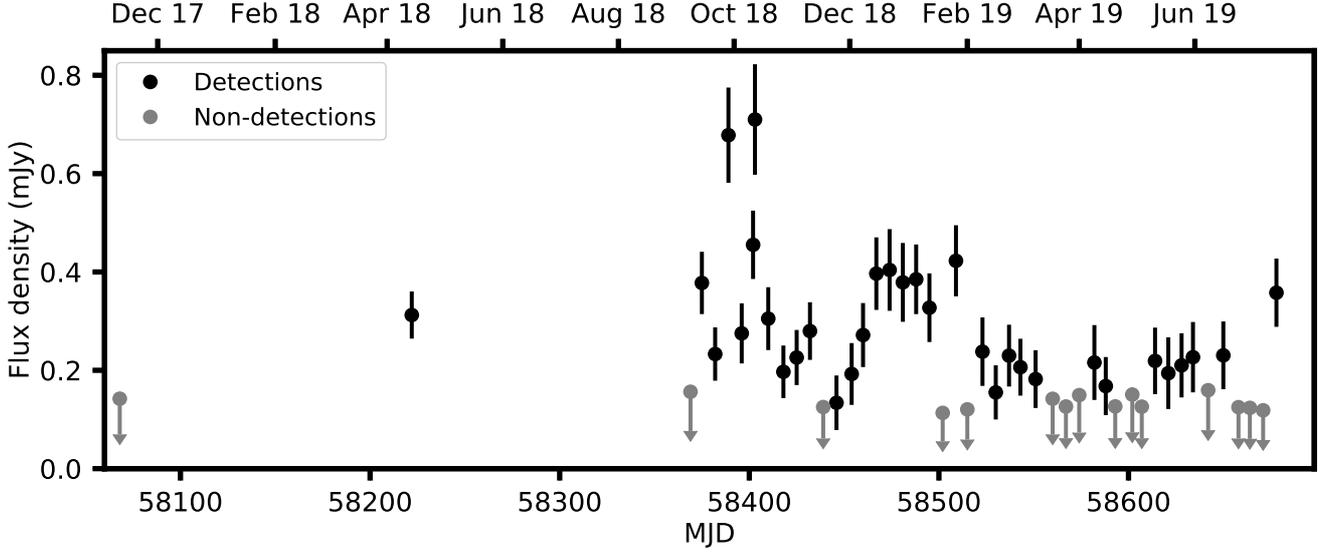}
 \caption{Radio light curve of \fb\, from the ThunderKAT observations described in Section\,\ref{sec: meerkat obs}. 
 We used a detection threshold of $2\sigma$, where $\sigma$ is the RMS noise close to the source.
 The upper limits are non-detections showing the $1\sigma$ RMS noise of the image local to the source.}
 \label{fig: Radio variability}
\end{figure*}

\begin{figure*}
 \includegraphics[width=\textwidth]{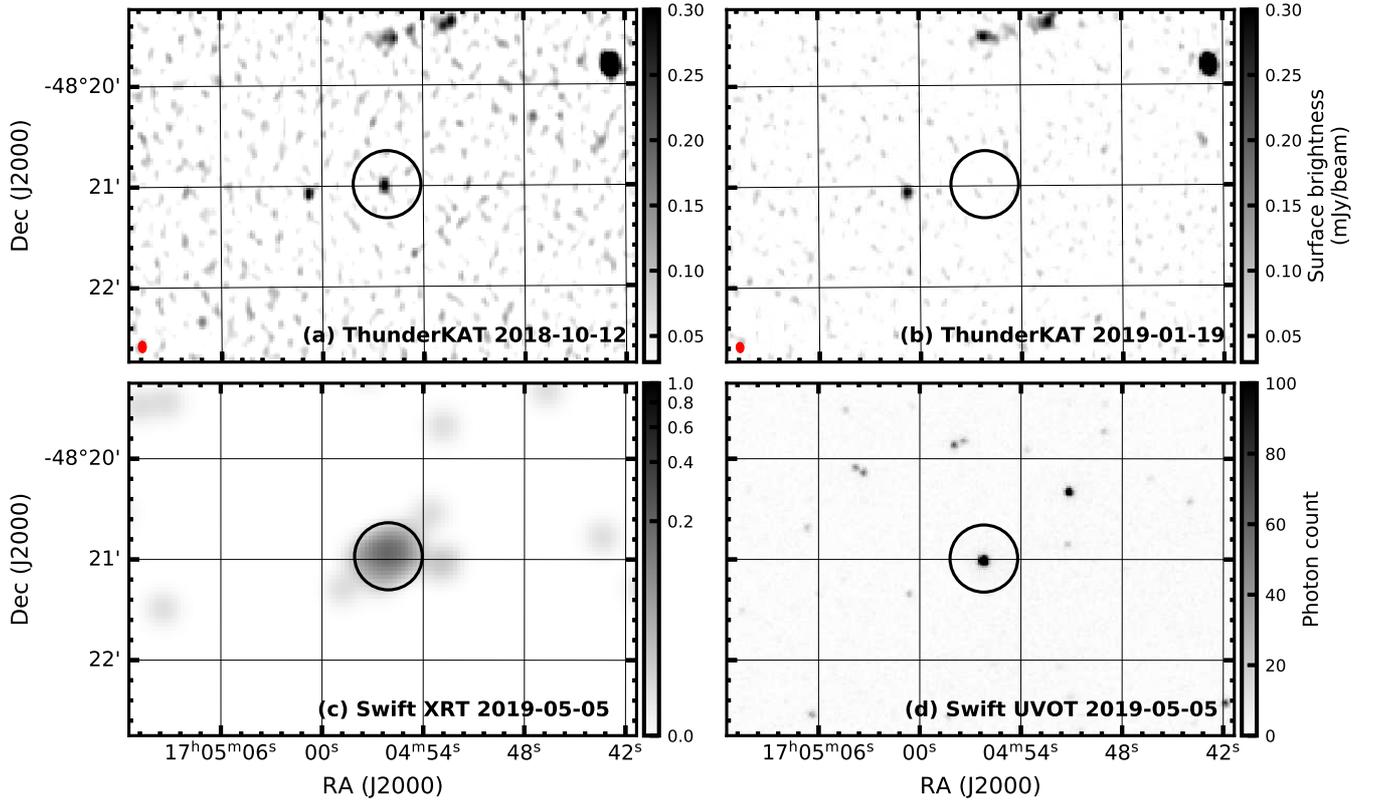}
 \caption{Radio, X-ray and UV images with \fb\, circled. Panels (a) and (b) show radio images from ThunderKAT. Panel (a) shows the source when it is detected at the highest flux density on 2018 Oct 12. Panel (b) shows an epoch, 2019 Jan 19, when the source is not detected. The synthesised beam is shown as a red ellipse in the bottom left corner of panels (a) and (b). Panels (c) and (d) respectively show the Swift XRT and Swift UVOT uvw1 images of the source from 2019 May 05. The grey scale on panels (a) and (b) shows the surface brightness in $\mathrm{mJy/beam}$, the grey scale in panels (c) and (d) shows photon count.}
 \label{fig: radio xray uv images}
\end{figure*}

Following the successful identification of \fb\, using \trap, the source was monitored in each weekly image of the \gx\, field.
The light curve of \fb, extracted using the \trap\, default parameters, for all 48 epochs is shown in Figure\,\ref{fig: Radio variability}.
Examples of MeerKAT images of the source are shown in Figure\,\ref{fig: radio xray uv images}.
To produce this light curve we forced \trap\, to take measurements at the position of \fb\, in every epoch, which means that \trap\, measured values even when the source was not detected, using a detection threshold of $2\sigma$ above the noise in the image local to the source.
In Figure\,\ref{fig: Radio variability} we plot these non-detections (which were confirmed by eye) as $1\sigma$ upper-limits where $\sigma$ is the RMS of the noise close to \fb.
The uncertainties on the flux density are the $1\sigma$ uncertainties calculated by TraP. The position of \fb\, places it at approximately the half power point of the MeerKAT primary beam when \gx\, is at the phase centre. On 2019 May 18 we observed \gx\, as usual, and also took an observation with \fb\, at the phase centre. Both observations were processed and calibrated in the same way. We then measured the flux density of the closest constant source to \fb\, in both images, for ease we will call this nearby source 170500$-$482103. 
We determined that the flux density of 170500$-$482103 when \fb\, is at the phase centre of the observation is $2.00\pm0.20$ times the flux density of 170500$-$482103 when \gx\, is at the phase centre. This indicates that 170500$-$482103 and \fb\, are at approximately the 50\,per\,cent point of the primary beam. Therefore, the flux density of \fb\, in this manuscript is the measured flux density multiplied by $2.00\pm0.20$ to correct for the primary beam effect.
Table\,\ref{tab: radio flux measurements} in Appendix\,\ref{App: flux measurements} includes the flux density measurements of the peak flux density of the source as measured by 
\trap, as well as the RMS noise measured near \fb.

We investigated shorter timescale variability by dividing each epoch of observations into $160\,\mathrm{second}$ images and running these shorter integration images through \trap. There is no variability in the 160 second integration images above one standard deviation from the mean of the light curve.

\section{Identification of the optical counterpart to MKT~J170456.2$-$482100}
\label{sec: optical ID}

The Python Blob Detector and Source Finder \citep[\texttt{pyBDSF};][]{2015ascl.soft02007M} was used for improved astrometry to determine the position of \fb\, in the epochs that it is detected. The position of the source was taken to be the mean of the J2000 positions: 17h04m56.2s $-$48d21m0.45s ($256.23450^{\circ}\,-48.35012^{\circ}$) with standard deviations of 1.2$\arcsec$ and 0.68$\arcsec$ for the Right Ascension and Declination respectively.
We then searched for sources nearby to this position and found the star \kstar, which has a \emph{Gaia} data release 2 \citep{2016A&A...595A...1G,2018AJ....156...58B,2018A&A...616A...1G} ICRS J2015.5 position of 17h04m56.25s $-$48d21m00.67s with uncertainties of $0.044\arcsec$ and $0.037\arcsec$ respectively, a proper motion of $-3.694 \times 10^{-3}\arcsec\,\mathrm{per\,year}$ and $-6.963 \times 10^{-3}\arcsec\,\mathrm{per\,year}$ in Right Ascension and Declination respectively, and a distance of $555\,\mathrm{pc}$ (68 per cent confidence interval 541--568\,pc). Taking proper motion into account, this position is $0.37\arcsec$ from the position of \fb, which means that we identify \kstar\, as the optical counterpart to \fb.

\kstar\, has been observed by numerous sky-surveys at various wavelengths. These include the original \emph{Hipparcos} Tycho-2 catalogue \citep{2000A&A...355L..27H}, the Gunn $I$ photometry from the Deep Near-Infrared Survey of the southern sky \citep[DENIS;][]{2000A&AS..141..313F}, the Two-Micron All-Sky Survey \citep[2MASS;][]{2003tmc..book.....C}, and the final \emph{Wide-Field Infrared Survey Explorer} catalogue \citep[\emph{AllWISE}][]{2013wise.rept....1C}.
\kstar\, was classified as a main-sequence K-dwarf (K7V) by \citet{2010PASP..122.1437P}, and it was classified by the All-sky Automated Survey \citep[ASAS;][]{1997AcA....47..467P} Catalogue of Variable Stars \citep[ACVS;][]{2002AcA....52..397P} as a first overtone delta Cepheid (DCEP$_{\mathrm{FO}}$) or eclipsing binary with a period of $0.955\,\mathrm{days}$. \citet{2012ApJS..203...32R} found, with a probability of $\sim50$ per cent, that the star is a type A Small Amplitude Red Giant \citep[SARG;][]{2004MNRAS.349.1059W} with a period of $21.24595\,\mathrm{days}$.

\section{Optical photometry}
\label{sec: optical photometry}

\kstar\, has been observed $\gtrsim4800$ times in total over the last $\sim$18 years by ASAS \citep{1997AcA....47..467P}, the Kilodegree Extremely Little Telescope \citep[KELT;][]{2007PASP..119..923P}, and the All-Sky Automated Survey for Supernovae \citep[ASAS-SN;][]{2014ApJ...788...48S,2017PASP..129j4502K}. The details of the photometric observations of \kstar\, made by these surveys are shown in Table\,\ref{tab: optical photometry}.

\begin{table*}
\caption{Summary of the archival optical photometry observations of \kstar.}\label{tab: optical photometry}
\centering
\begin{tabular}{p{1.5cm}p{1.0cm}p{1.8cm}p{2.0cm}p{1.8cm}p{1.5cm}}
    \hline
    Survey & Optical band & Date of first observation & Date of last observation & Number of observations & Number of semesters \\[1mm]
    \hline
ASAS    & V & 2001 Jan 01   & 2009 Oct 12   & 534   & 9 \\
KELT    & V & 2013 May 15   & 2015 Oct 13   & 3037  & 3 \\
ASAS-SN & V & 2016 Mar 10   & 2018 Sep 22   & 762   & 5 \\
ASAS-SN & g & 2018 Feb 26   & Ongoing       & 645   & 3 \\

\hline
\end{tabular}
\end{table*}

Once \kstar\, was identified as the source of \fb, we observed the star with the HIPPO Photopolarimeter \citep{Potter2010} on the 1.9-m telescope at the South African Astronomical Observatory (SAAO), Sutherland, South Africa. HIPPO uses photomuliplier detectors, so is a photon counting instrument, obtaining measurements simultaneously in two channels. Dual counter-rotating waveplates (1/2 and 1/4) modulate the signal at 10 Hz and allow simultaneous determination of all four Stokes parameters. The intrinsic time resolution is 1\,ms for total intensity measurements (I) and 100\,ms for the three other Stokes parameters (Q, U \& V). In practice data are accumulated over many 100 ms cycles, typically several 100\,s, until a sufficient S/N is obtained. A log of these photopolarimetric observations is presented in Table~\ref{oplog}.

\begin{table*}
\centering
\caption{Optical observation log for \kstar, including the radial velocity measurements.}\label{oplog}
\begin{tabular}{p{1.65cm}|l|p{1.8cm}|l|l|p{1.2cm}|p{1.5cm}|p{1.95cm}}
  \hline
Telescope & Instrument & Wavelength range/Filter (\AA) &  Resolving Power & Date & UT Start (hh:mm) &  Exposure times & Radial velocity ($v_{\rm LSR}$, km s$^{-1}$) \\[1mm]
   \hline
SAAO 1.9-m  & HIPPO &3400$-$9000; no filter &  &  2019-03-12 & 02:19 & 467s & \\
 &  & B & &  & 02:46 & 2020s & \\
 &  & V &  &  & 02:48 & 2073s & \\
 &  & R &  &  & 02:50 & 2065s & \\
  &  & I &  &  & 02:51 & 1987s & \\
SALT    & HRS & 3800$-$8800 & 15000 & 2019-02-25 & 02:42 & 1200s & 53 $\pm$ 3 \\
              &         &  &  & 2019-03-24 & 00:16 & 1500s & 0 $\pm$ 3\\
              &         &  &  & 2019-04-17 & 22:45 & 1500s & --30 $\pm$ 3\\
              &         &  &  & 2019-04-18 & 22:38 & 1500s & --32 $\pm$ 3\\
              &         &  &  & 2019-06-08 & 19:18 & 1500s & 38 $\pm$ 3\\
            &         &  &  & 2019-06-10 & 19:30 & 1500s & 52 $\pm$ 3\\
SALT    & RSS & 6137$-$6953 & 4350 & 2019-03-01 & 01:55 & 600 $\times$ 5s & 20 $\pm$ 18\\
Cerro Tololo LCO    & NRES & 3800$-$8600    & 53000 & 2019-06-15 & 19:04 & 1200s & 24 $\pm$ 7 \\
Sutherland LCO      & NRES & 3800$-$8600    & 53000 & 2019-06-01 & 01:20 & 1200s & --35 $\pm$ 7 \\
                    &      &                &       & 2019-06-04 & 03:05 & 1200s & --21 $\pm$ 7 \\
                    &      &                &       & 2019-06-06 & 02:08 & 1200s & 3 $\pm$ 7 \\
                    &      &                &       & 2019-06-09 & 03:34 & 1200s & 37 $\pm$ 7\\
                    &      &                &       & 2019-06-14 & 03:06 & 1200s & 43 $\pm$ 7 \\
                    &      &                &       & 2019-06-19 & 07:20 & 1200s & --16 $\pm$ 7\\
                    &      &                &       & 2019-07-04 & 03:33 & 1200 & 49 $\pm$ 7\\
                    &      &                &       & 2019-07-05 & 23:08 & 1200 & 43 $\pm$ 7\\
\hline
\end{tabular}
\end{table*}

\subsection{Photometric variability}

The automated ASAS pipeline classified \kstar\, as a 0.955043 day variable. However, we performed a Lomb-Scargle \citep{1976Ap&SS..39..447L, 1982ApJ...263..835S,2018MNRAS.478..218P} analysis of the data and found that the 0.955043 day periodicity is due to the underlying periodicity of the observing cadence, instead of the intrinsic periodicity of the star. We found the highest precision period of the star to be $21.25\pm0.04\,\mathrm{days}$ by performing a Lomb-Scargle analysis on the combined V-band data for the full $\sim$18 years of observations. This is consistent with the period of $21.246\,\mathrm{days}$ found by \citet{2012ApJS..203...32R} using machine learning techniques on the ASAS observations.

\begin{figure}
 \includegraphics[width=\columnwidth]{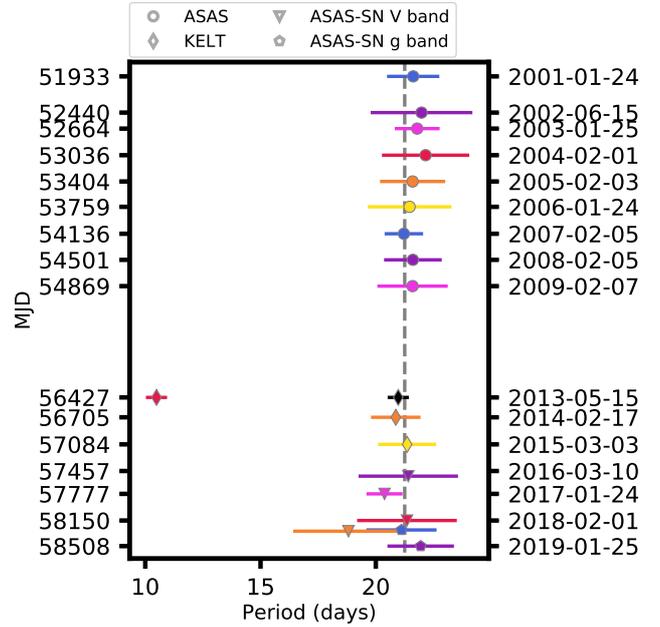}
 \caption{Periods found using the Lomb-Scargle algorithm for the different data sets divided into semesters, the MJD and date shown on the y axis is the date of the first observation in the semester. The semesters are shown chronologically from top to bottom. The grey dashed line is the period found by performing a Lomb-Scargle on all of the V band observations together: $P=21.25\pm0.04\,\mathrm{days}$. In the first semester of KELT observations (start date: 2013 May 15) there was more power in the second harmonic of the period.}
 \label{fig: Optical Periods}
\end{figure}

\begin{figure*}
 \includegraphics[width=0.85\textwidth]{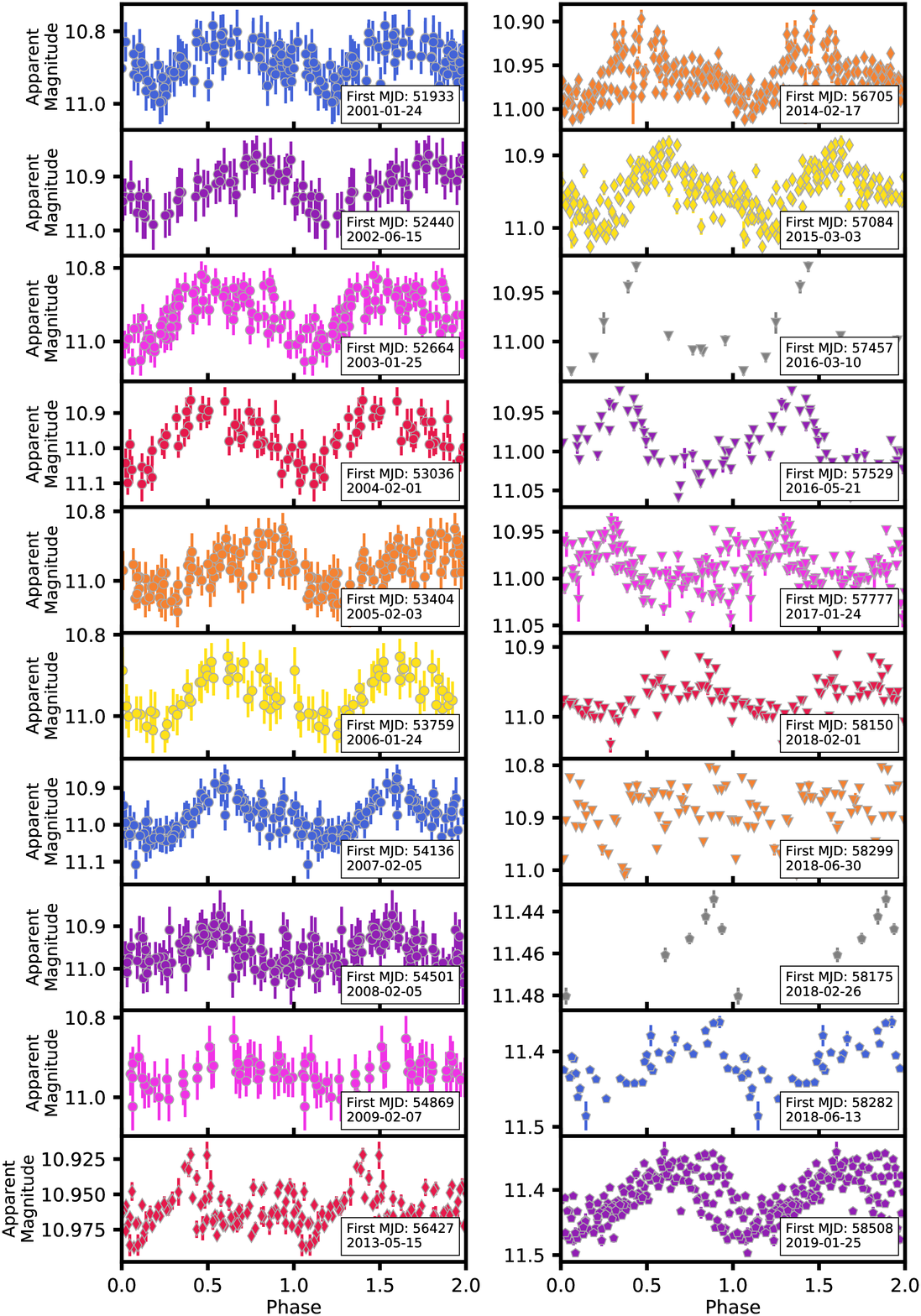}
 \caption{Optical light curves folded to a period of $21.25\pm0.04\,\mathrm{days}$. The observations are divided into semesters and have been folded and phase corrected to MJD=53571.1. The MJD of the first observation of the semester is shown in the bottom right of each panel. The colours and symbols are the same as in Figure\,\ref{fig: Optical Periods}, the grey points are the semesters that did not have enough points to perform a Lomb-Scargle analysis. The observations have been re-binned such that there is one measurement per day and outliers have been removed.}
 \label{fig: Folded Light Curves}
\end{figure*}

We further investigated the period of \kstar\, by performing a Lomb-Scargle analysis on each semester\footnote{We define a semester as the period in each year when the source is a night time source.} of observations from ASAS, KELT, and ASAS-SN. We found that the period for each semester is consistent with $21.25\pm0.04\,\mathrm{days}$, as shown in Figure\,\ref{fig: Optical Periods}.
While the period over $\sim18\,\mathrm{years}$ remains stable, Figure\,\ref{fig: Folded Light Curves} shows that the shape of the folded light curve and the phase of the peak varies over time. In some semesters the folded light curves show little or no variability and the period is less distinct. In the first semester of KELT observations (start MJD=56427, 2013-03-15) the Lomb-Scargle analysis results in more power at half of the $21.25\pm0.04\,\mathrm{days}$ period. Figure\,\ref{fig: Folded Light Curves} shows that this is due to a second peak in the folded light curve of this semester.

\subsection{Spectral energy distribution}
\label{SEDsect}

\begin{figure}
 \includegraphics[width=\columnwidth]{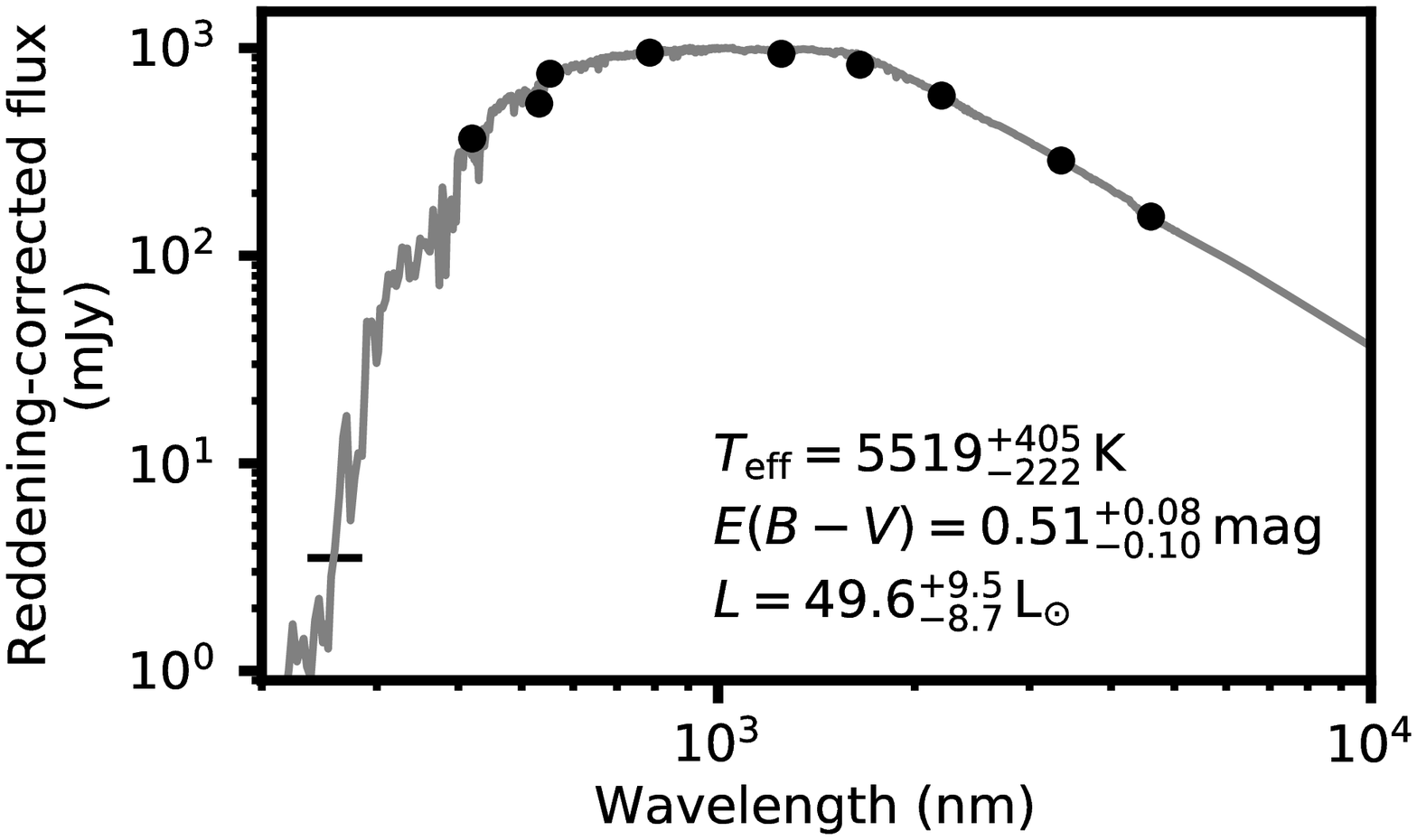}
 \caption{Spectral energy distribution of \kstar, showing literature sources used in the fit (black data points, see text). The black line shows our Swift UVOT observation, with length of the line denoting the width of the filter. The grey spectrum shows the best-fit {\sc bt-settl} model, with the parameters shown.}
 \label{SED plot}
\end{figure}

To obtain fundamental stellar parameters, the optical and infrared spectral energy distribution (SED) of \kstar\, was modelled. The SED was constructed from optical $B_{\rm T}$- and $V_{\rm T}$-band fluxes, obtained from the original \emph{Hipparcos} Tycho-2 catalogue \citep{2000A&A...355L..27H}; the mean ASAS $V$-band magnitude \citep[][]{1997AcA....47..467P}; Gunn $I$ photometry from DENIS \citep[][]{2000A&AS..141..313F}; $J$, $H$ and $K_{\rm s}$ magnitudes from 2MASS \citep[][]{2003tmc..book.....C}; and the [3.4] and [4.6] magnitudes from \emph{AllWISE} \citep[][]{2013wise.rept....1C}. The [11.3] and [22] magnitudes from \emph{AllWISE} were not used, as \kstar\, lies 59${\arcsec}$ from IRAS 17011-4817, an optically obscured star that reaches second magnitude in these bands and provides considerable contaminating flux. 

The software described in \citet{2012MNRAS.427..343M,2017MNRAS.471..770M} was used to fit {\sc bt-settl} model atmospheres to the star. The scattering law of \citet{2003ApJ...598.1017D}, assuming $R_{\rm V} = E(B-V) / A_{\rm V} = 3.1$, was used to correct the observed photometry for the effects of interstellar reddening, with $E(B-V)$ as a free parameter. The \emph{Gaia} Data Release 2 distance of 555 pc from \citet{2018AJ....156...58B} was adopted to scale the luminosity. A mass of 2.5 M$_\odot$ was assumed to set the model surface gravity (log($g$)), but this has relatively little impact on the results.

Assuming a single stellar light source, a best fit was found for a temperature of $T_{\rm eff} = 5519 ^{+405}_{-222}$ K and $E(B-V) = 0.51 ^{+0.08}_{-0.10}$ mag, giving a luminosity scaling of $L = 49.6 ^{+9.5}_{-8.7}$ L$_\odot$. Unaccounted-for uncertainties include any intrinsic variability, contamination from other sources such as unresolved companions, or other departures from a simple model atmosphere, and a degree of correlation exists among the three parameters. 
The reduced $\chi^2$ per degree of freedom is 0.93, hence the expected impact of these uncertainties is small.

To determine physical parameters from these values, stellar isochrones were generated using the Padova stellar evolution code\footnote{\href{http://stev.oapd.inaf.it/cgi-bin/cmd\_3.2}{http://stev.oapd.inaf.it/cgi-bin/cmd\_3.2}} \citep{2017ApJ...835...77M}. The default settings were used, including the assumption of solar metallicity ($Z_\odot = 0.0152$). The star was identified as most likely being a sub-giant, completing its transition across the Hertzsprung gap to the base of the red giant branch. This period is characterised by a rapid inflation of the star towards the Hayashi limit as it becomes convective, following exhaustion of core hydrogen. Without a detailed metallicity derivation, a precise age and mass is impossible to ascertain. However, the age of the star is consistent with solar-metallicity isochrones in the age range $650 \lesssim t \lesssim 870$ Myr, corresponding to a stellar mass of $2.22 \lesssim M \lesssim 2.48$ M$_\odot$ and $\log(g) \approx 3.03 \pm 0.15$ dex. These values assume that the flux from any possible companion is negligible over the optical and near-IR SED, and that the star has taken an evolutionary path through the H--R diagram comparable to a single star with similar initial parameters.

\subsection{Photopolarimetry}

\kstar\, was observed using HIPPO on 2019 March 12.
The results show that \kstar\ was not varying significantly photometrically over the $\sim$ 2 h observation period and furthermore show no evidence for higher frequency variability over timescales of $\sim$ 1 $-$ 1000 s. An initial filterless observation was performed followed by five repeat BVRI filter sequences. There was no evidence of variations in the polarization between different filter sequences and the average polarization parameters for the BVRI filters are presented in Table~\ref{pol}.
The level of linear polarization measured is consistent with the values expected for the ISM at the Galactic coordinates for \kstar.

\begin{table}
\caption{Polarization measurements for \kstar}
\centering
\begin{tabular}{l|l|l|l|r}
  \hline
Filter & Linear ($\%$) &  $\theta$ ($^{\circ}$) & Circular ($\%$)  \\[1mm]
   \hline
B & 2.56 $\pm$	0.32 &	39.72 $\pm$	4.44 &	$-$0.24 $\pm$	0.22\\
V & 2.87 $\pm$	0.18 &	36.78 $\pm$	2.93 &	0.00 $\pm$	0.43\\
R & 2.61 $\pm$	0.08 &	39.21 $\pm$	2.30 &	$-$0.04 $\pm$	0.09\\
I & 2.33 $\pm$	0.16 &	41.11 $\pm$	1.59 &	0.01 $\pm$	0.12\\
\hline
\end{tabular}
\label{pol}
\end{table}

\section{Optical spectroscopy}
\label{sec: optical spectroscopy}

The variable nature of the optical light curve of \kstar\, is suggestive of an active flare star or binary companion. We therefore undertook a spectroscopic campaign to further understand the nature of the variability.
We observed \kstar\, with the Southern African Large Telescope \citep[SALT;][]{Buckley2006} situated at the SAAO using the High Resolution Spectrograph \citep[HRS;][]{Bramall2012,Crause2014} and the Robert Stobie Spectrograph \citep[RSS;][]{Burgh2003}. HRS is a fibre-fed, dual-beam, white pupil, vacuum-stabilised high resolution (R = 15,000$-$80,000, depending on mode and wavelength) spectrograph, while RSS is a prime focus low to medium resolution (R = 350$-$6500, depending on grating/angle/slit choice) slit spectrograph.
\kstar\, was also observed by the Las Cumbres Observatory \citep[LCO;][]{2013PASP..125.1031B} Network of Robotic Spectrographs (NRES). The NRES network uses high-resolution (R$\sim$53000) optical echelle spectrographs on two, $1\,\mathrm{m}$ telescopes simultaneously with a ThAr calibration source.
The details of the spectroscopic observations are summarised in Table\,\ref{oplog}.

\subsection{High-resolution spectroscopy}

Spectra of \kstar\ were taken with the SALT HRS in the Low Resolution (LR) mode on six occasions, shown in Table\,\ref{oplog}, in clear conditions with $\sim1\arcsec$ seeing.  Both blue (3800$-$5550\AA) and red (5450$-$9000\AA) spectra were obtained and reduced using the weekly set of HRS calibrations, including ThAr arc spectra and QTH lamp flat-fields. 

The primary reduction was conducted using the SALT science pipeline, PySALT \citep{Crawford2016}, correcting for overscan, bias and gain correction. The spectral reductions were then undertaken using a MIDAS-based echelle reductions package \citep[see details in][]{Kniazev2016}. For the blue spectra, the average S/N had a minimum of $\sim$10 for the bluest orders ($\sim$3980$-$4000\AA), which increased linearly to $\sim$120 at 4600\AA, and reached a maximum of $\sim$140 at 5400\AA. For the red spectra, the S/N increased linearly from $\sim$80 at 5550\AA\, to $\sim$180 at 7000\AA, peaking at $\sim$200 by 8000\AA\, and then declining slightly to $\sim$160 by 8800\AA. This behaviour is consistent with the late spectral type (K4-5) attributed to \kstar.

\subsection{Fast time resolved spectroscopy}

\begin{figure}
 \includegraphics[width=\columnwidth]{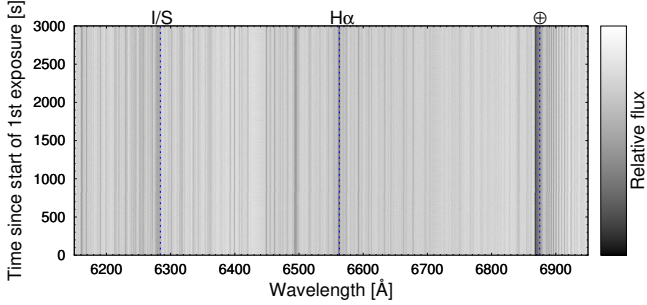}
 \caption{$600\times5$\,s SALT RSS red spectra, covering 6137$-$6953\AA, of \kstar\, taken on 1 March 2019. The absorption lines include the narrow H$\alpha$ line (right of centre), the telluric oxygen B-band (far right, marked with $\bigoplus$) and many other photospheric lines. Each spectrum was normalized to account for slit losses.}
 \label{trailed-spectra}
\end{figure}

High time resolution spectroscopy of \kstar\ was obtained on 1 March 2019 using the RSS in frame transfer mode in clear conditions with $1\arcsec$ seeing. The PG2300 grating was used at a grating angle of 49.25$^{\circ}$, which gave a wavelength coverage of 6137$-$6953\AA\ at a mean resolution, with a $1.5\arcsec$ slit, of 1.5\AA. A set of 600 continuous 5 second exposures were obtained, with no dead time between frames because of the frame transfer mode. The data were reduced using the PySALT package \citep{Crawford2016} to make the usual corrections for overscan, bias, gain and crosstalk. A 2\,s neon arc lamp exposure, taken after the observations, was used to wavelength calibrate the spectra. The IRAF package was then used to perform the spectral reductions, using the tasks in \texttt{twodspec}. In Figure~\ref{trailed-spectra} we show a 2D spectral image comprising of all 600 spectra, extending from 6137$-$6953\AA\ and stacked sequentially in time, increasing upwards. 

Little variation is seen in the spectra, in terms of line strength or position, and certainly no evidence of H$\alpha$ variability.

\subsection{Interpretation}

\subsubsection{Radial velocity curve}

\begin{figure}
 \includegraphics[width=\columnwidth]{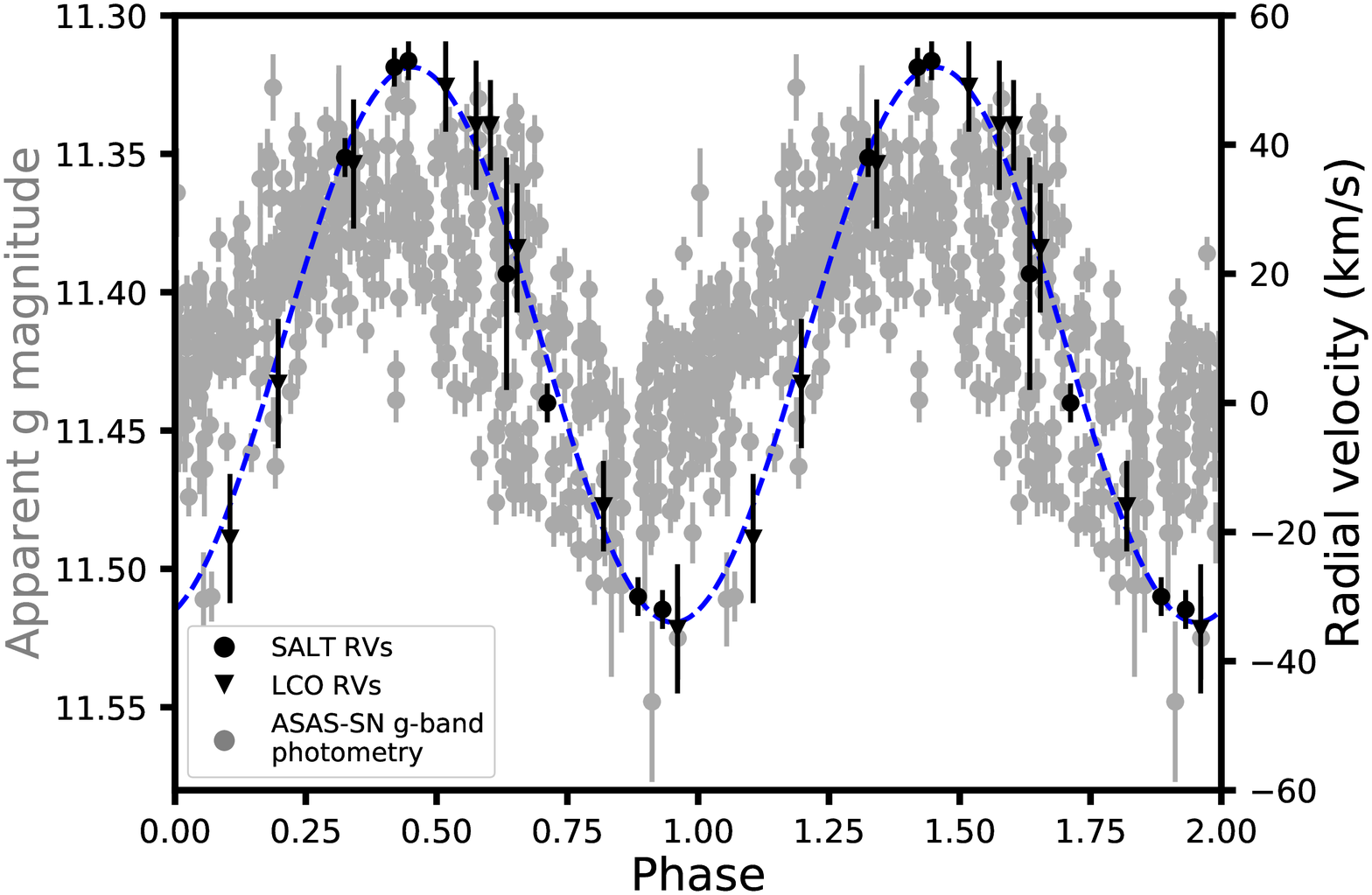}
 \caption{Radial velocity (black) and optical photometry (grey) curves, phased to a period of $21.25\pm0.04\,\mathrm{days}$ and an epoch of MJD=53571.1.
The blue dashed line represents a visua, sinusoidal fit to the radial velocity points.}
 \label{rvcurve}
\end{figure}

The wavelength calibration of the first SALT/HRS spectrum was initially checked against a telluric-feature spectrum in the regions around 6000 and 6800 \AA, and a good match was found. The radial velocity amplitude of the target star was then derived by cross-correlating the spectrum with a {\sc bt-settl} model atmosphere \citep[$T_{\rm eff} = 5000\,\mathrm{K},\,\log g = 4.0\,\mathrm{dex}$, solar metallicity;][]{2003IAUS..211..325A}
in the region 6100--6240 \AA, chosen to be largely free of telluric absorption or emission features. Radial velocities for subsequent spectra were checked in a similar manner, then cross-correlated against this first spectrum, to ensure a more accurate relative velocity. Radial velocities for the LCO spectra were corrected from the pipeline-reduced version by --88 km s$^{-1}$. Observed radial velocities were corrected to the Local Standard of Rest using the {\sc starlink} tool {\sc rv} \citep{2014ASPC..485..391C}. The resulting radial velocities are listed in Table \ref{oplog} and shown in Figure\,\ref{rvcurve}. Formal errors are dominated by the accuracy of the wavelength calibration, and approximate uncertainties have been assigned on this basis. Since these cross-instrument velocity uncertainties dominate the radial velocity amplitude, and since the photometric period is well known, we do not perform a formal fit to the radial velocity data.

\subsubsection{Spectral characterisation}
\label{sec: spectral char}

The spectrum of \kstar\, appears consistent with the K giant determined from the SED modelling in Section \ref{SEDsect}. The CaII H and K lines are in emission reaching the continuum level and there is infilling of the infrared triplet, which indicates that the star is chromospherically active (the Ca $\lambda$8662 line is shown in Figure \ref{specvar}).

There are no emission lines in the optical spectrum other than the chromospheric emission of the H and K lines and the infilling of the infrared triplet. This indicates a general lack of hot material in the vicinity of the unseen companion, as emission lines are generally seen even in relatively faint accretion discs \citep[e.g. RW Sextantis;][]{2017MNRAS.470.1960H}.

\subsubsection{Spectral variation}
\label{sec: spectral var}

\begin{figure*}
 \includegraphics[height=\columnwidth,angle=-90]{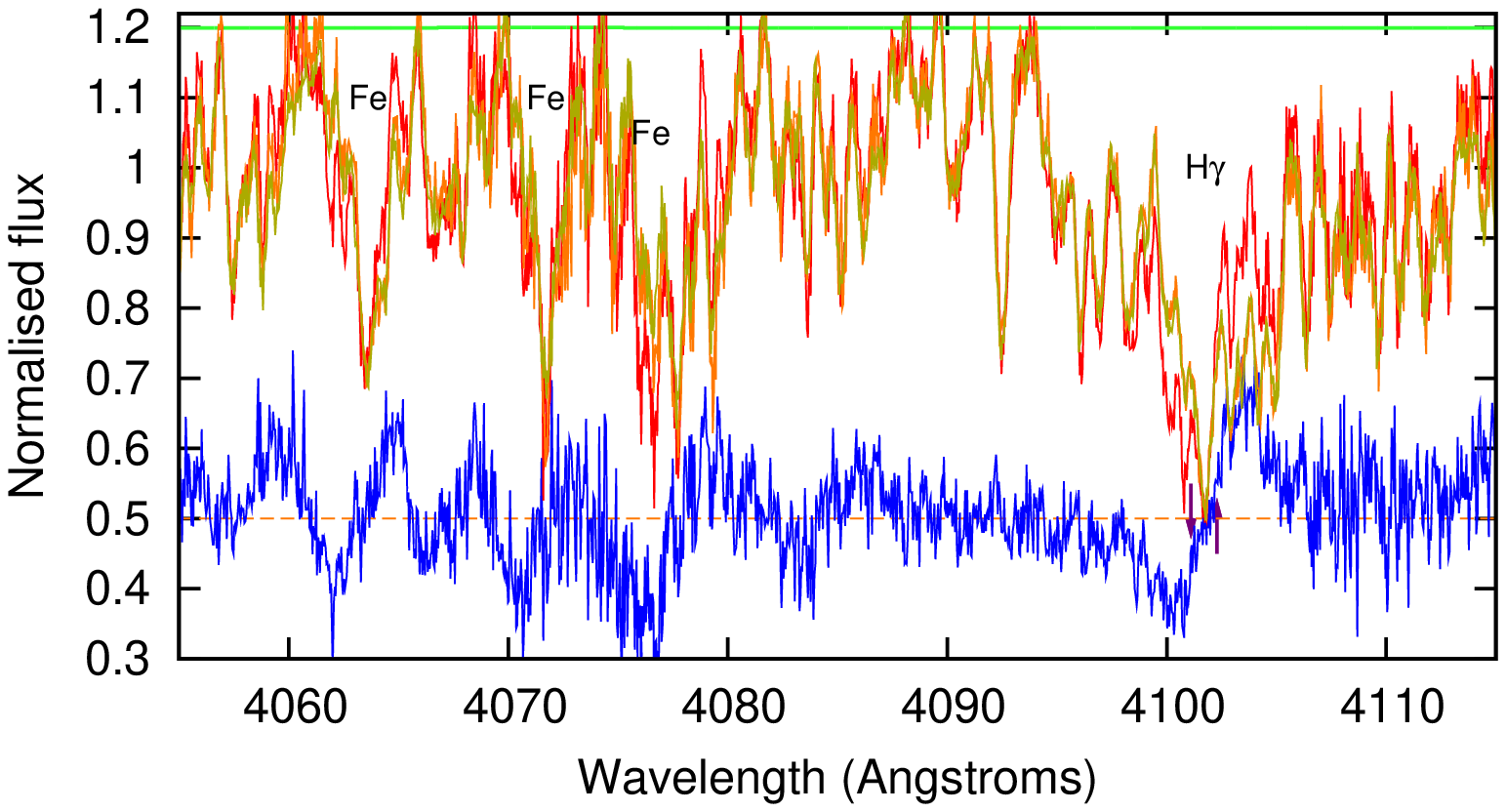}       
 \includegraphics[height=\columnwidth,angle=-90]{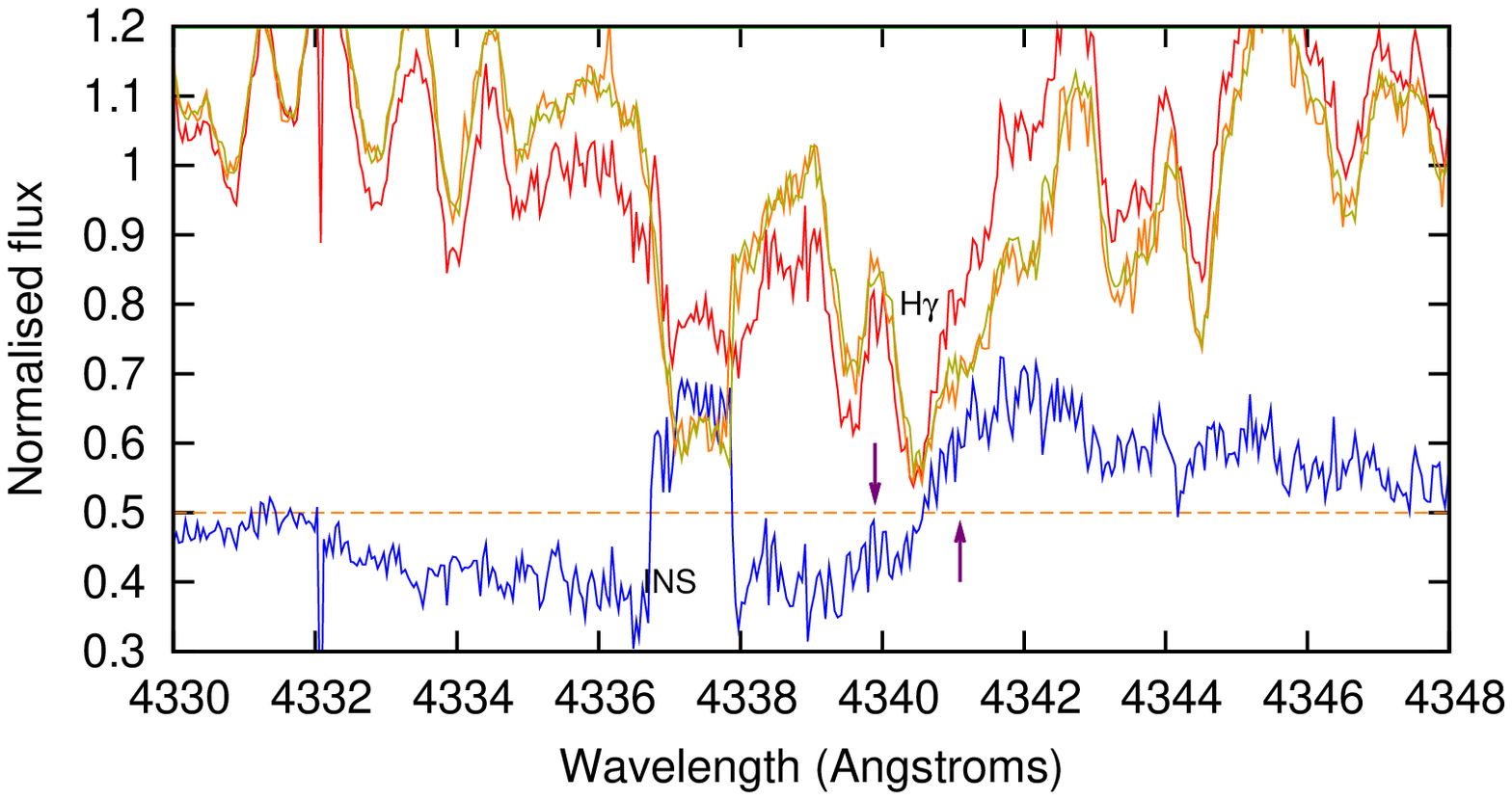}      
 \includegraphics[height=\columnwidth,angle=-90]{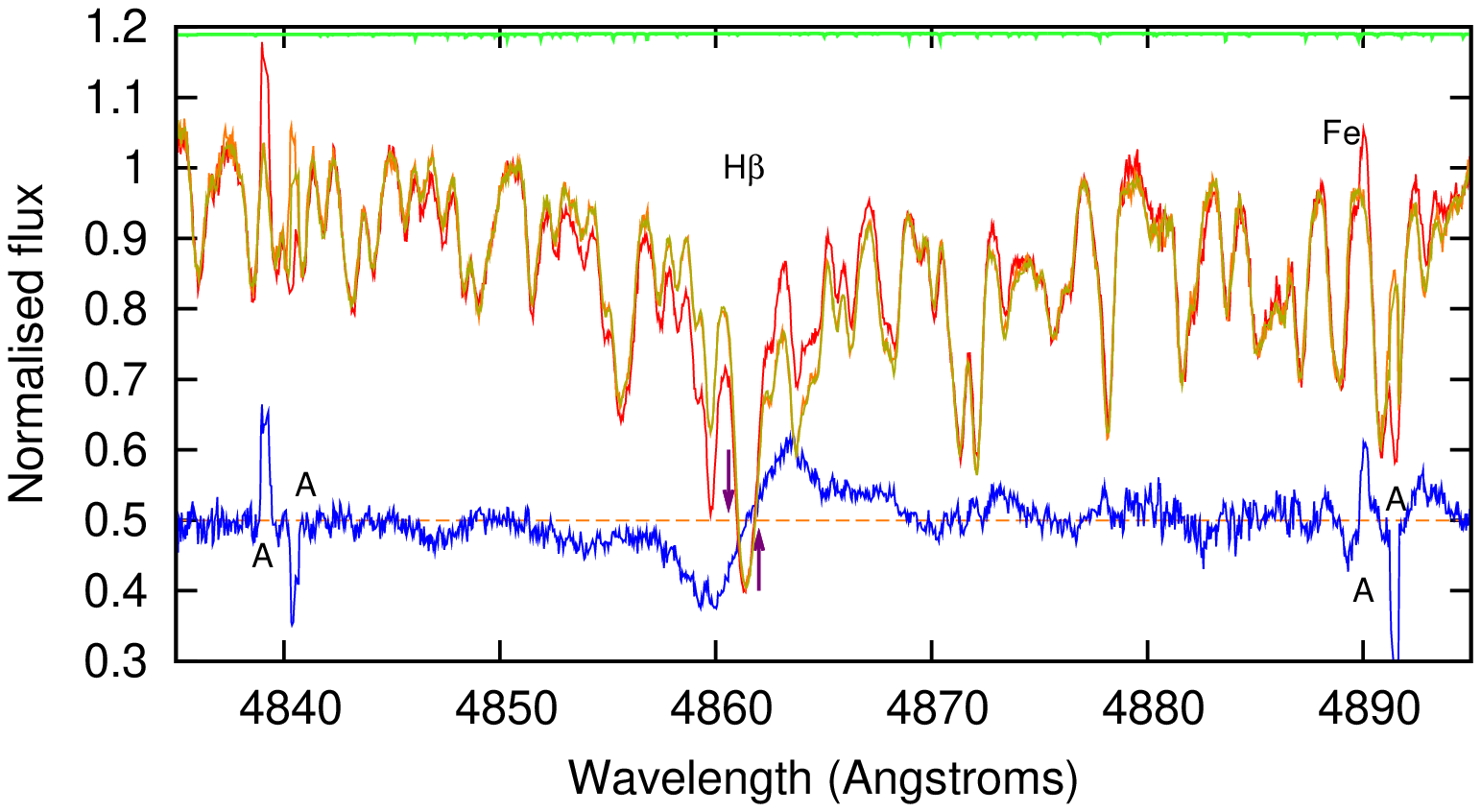}     
 \includegraphics[height=\columnwidth,angle=-90]{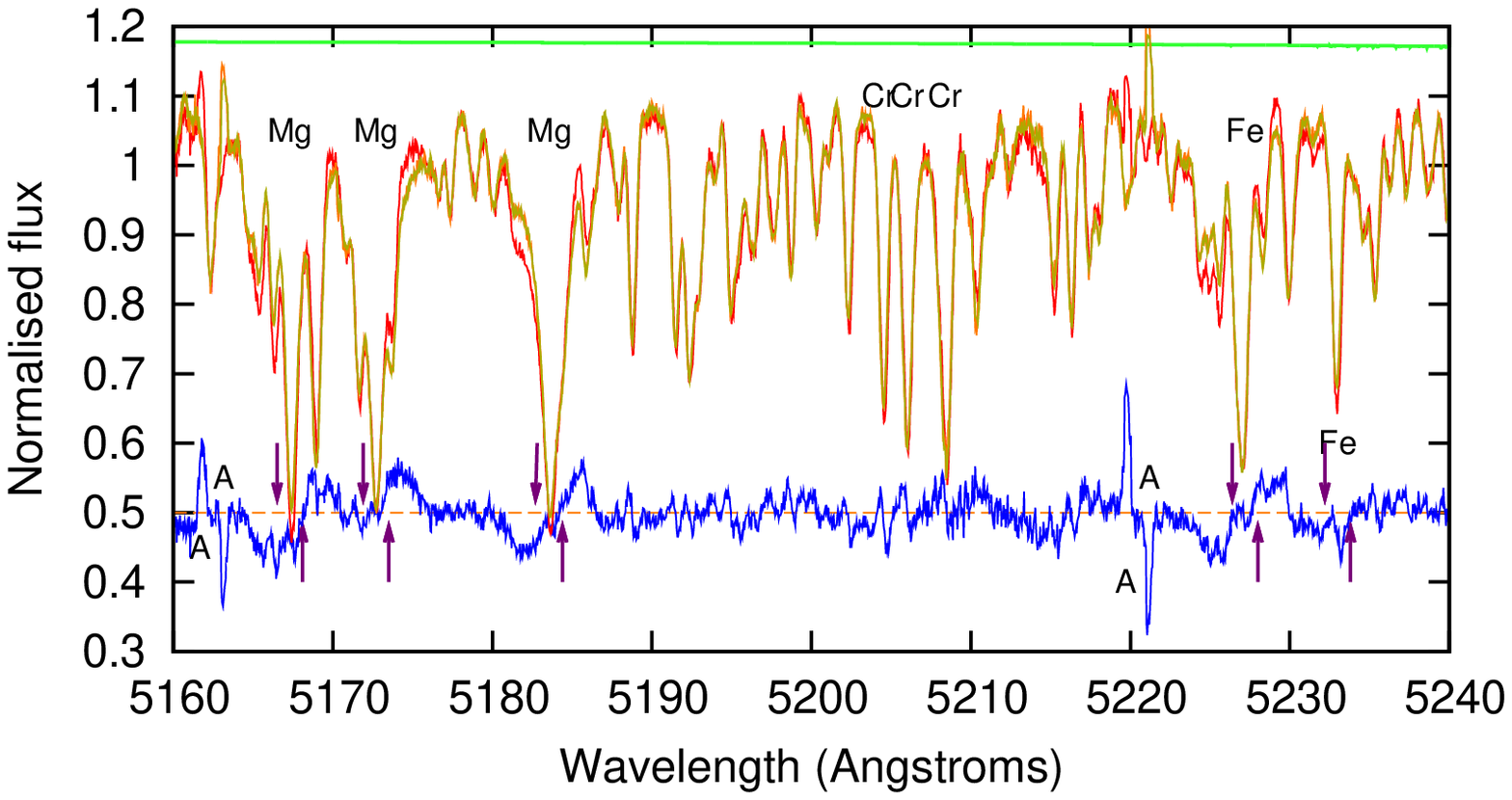}    
 \includegraphics[height=\columnwidth,angle=-90]{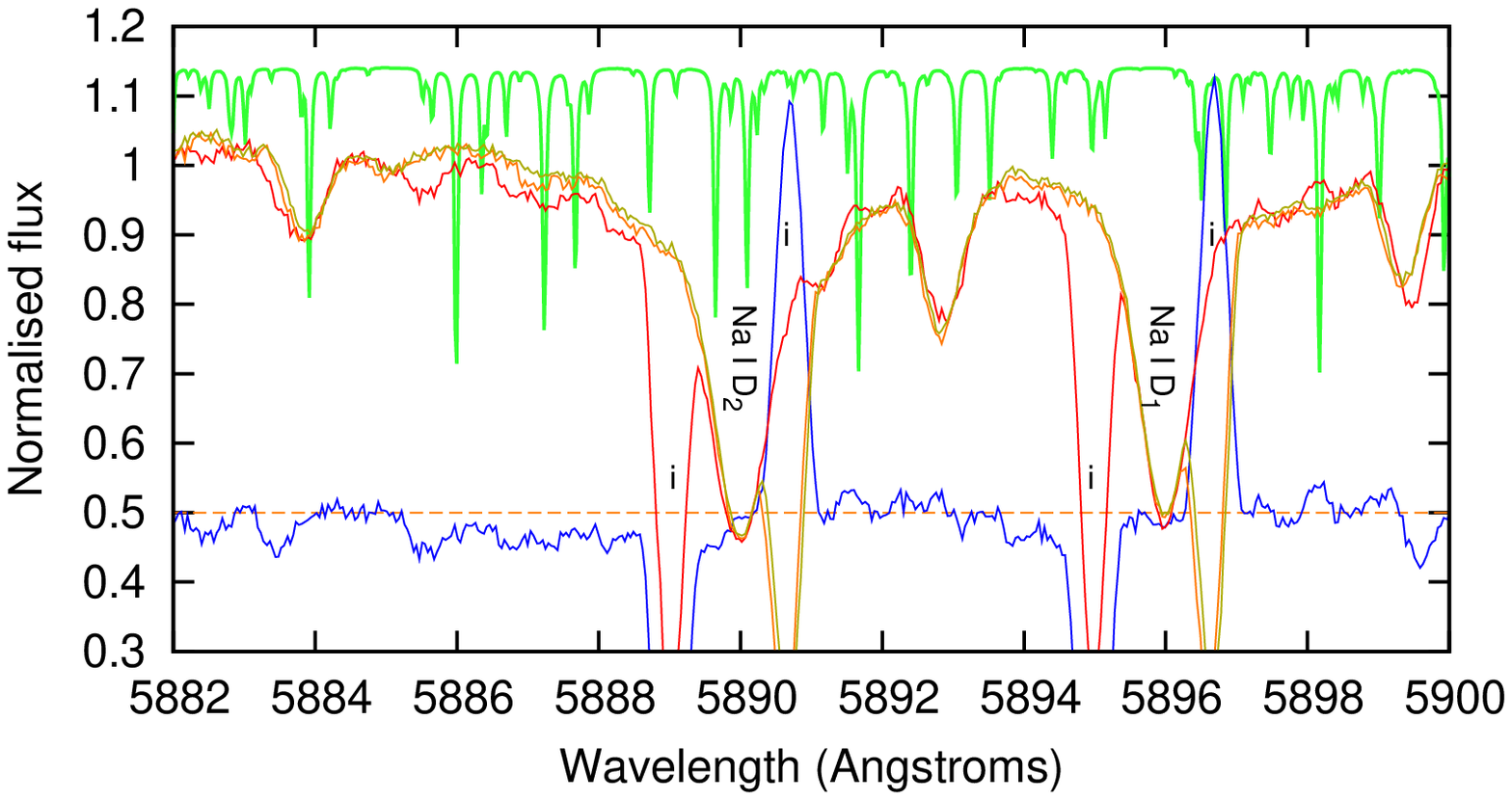}   
 \includegraphics[height=\columnwidth,angle=-90]{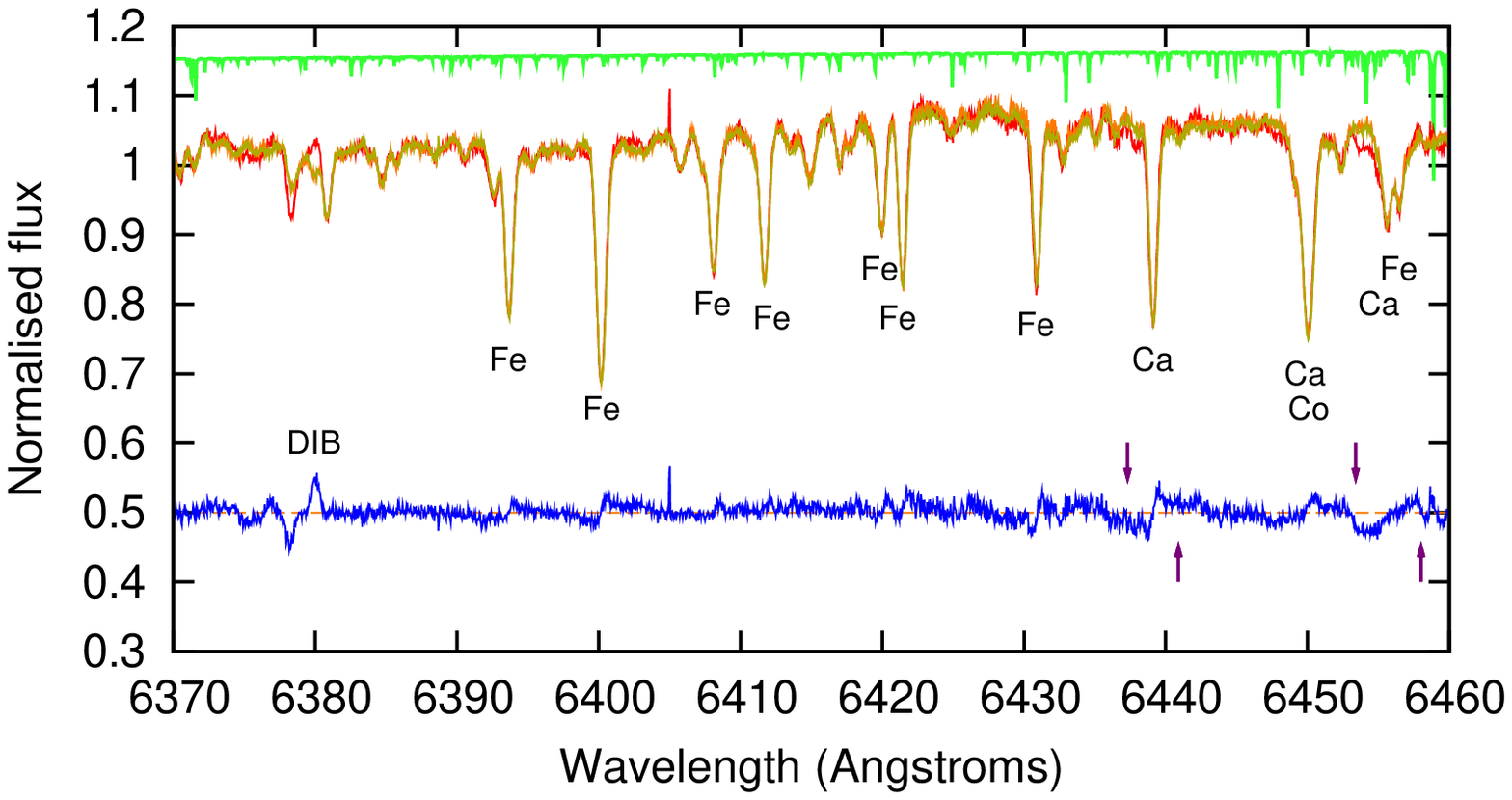}  
 \includegraphics[height=\columnwidth,angle=-90]{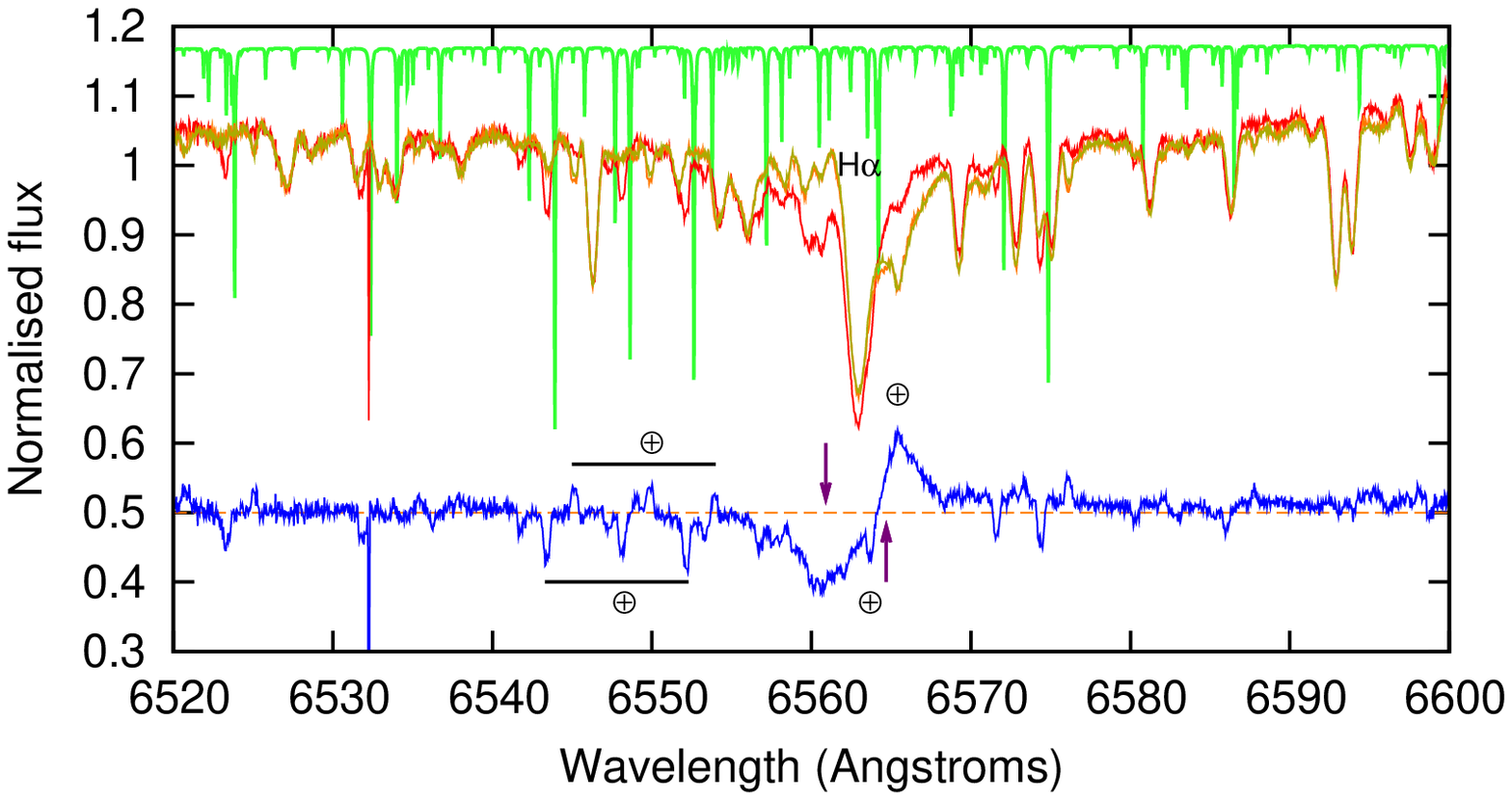}
 \includegraphics[height=\columnwidth,angle=-90]{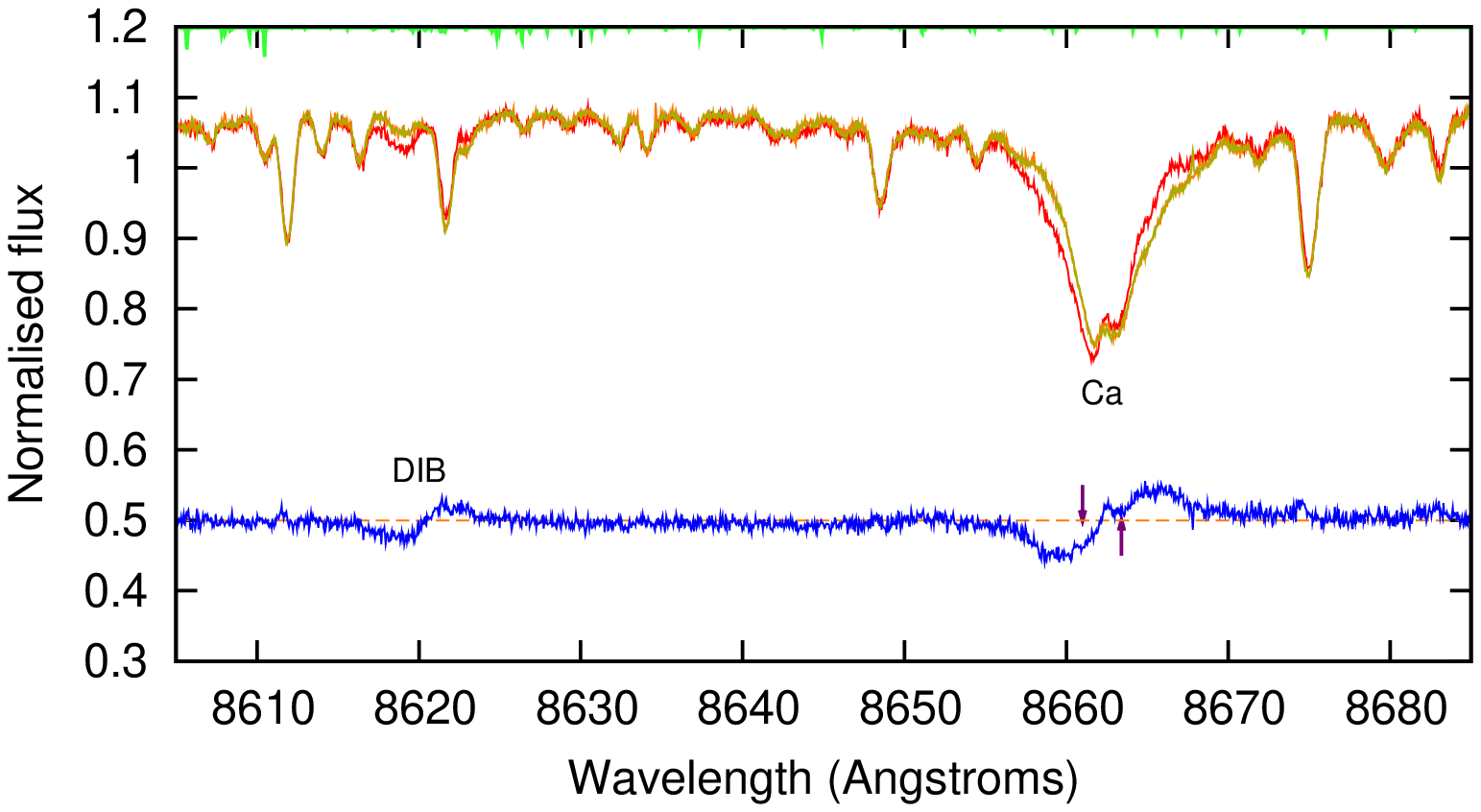}
\caption{SALT/HRS spectra, flux-normalised and corrected to the radial velocity amplitude of the giant star. The first, second and fourth SALT/HRS spectra from Table \ref{oplog} are shown in red, orange and gold. Of the spectra taken, these represent those with the most-positive, closest-to-average and most-negative radial velocities. The light green spectrum (top) shows the telluric absorption at rest velocity, offset vertically by +0.2. The difference between the first and fourth spectra is shown in blue (bottom), offset by +0.5 in normalised flux. Instrumental artefacts (INS), telluric absorption ($\oplus$) and airglow emission (A), and interstellar (i, DIB) lines are visible in this differenced spectrum. Purple arrows represent the systemic rest velocity and are explained in the text in Section\,\ref{sec: spectral var}. The minima/maxima of the blue difference spectrum are seen +/-$\sim$12 km s$^{-1}$ beyond these arrows.}
 \label{specvar}
\end{figure*}

Despite showing considerable radial velocity variations, the shape of the red portion of the \kstar's spectrum remains mostly constant, generally to within $\sim$1 per cent (Figure \ref{specvar}). Most variations appear to be caused by telluric features, interstellar gas lines and diffuse interstellar bands (DIBs). Subtle, percentage-level variations in the immediate vicinity of strong atomic lines can be traced to variation in line widths, likely due to very small variations in the recorded spectrum caused by either seeing or telescope focus.

However, variations in the line wings of bright lines exist. In the difference spectrum, this results in negative and positive features offset from the stellar velocity. The systemic rest velocity ($\sim \pm$43 km s$^{-1}$) is marked with purple arrows in Figure \ref{specvar}. In stronger and bluer lines, extra absorption is seen beyond the systemic rest velocity, indicating the absorbing material is on the opposite side of the system's barycentre and hotter than the giant star. Modelling these components as one would a companion star gives a reflex velocity amplitude of $K_2 \approx 12 \pm 10$ km s$^{-1}$. The relative line depths suggest a spectral temperature of order 7000 K, but this is very approximate as the K-giant star's spectrum cannot be accurately subtracted with our current data.

\section{Radio pulsation searches}
\label{sec: pulsation searches}

As \fb\, is coincident with a star in a spectroscopic binary, it is possible that the companion star may be a pulsar or pulsating white dwarf, similar to AR Scorpii~\citep{2016Natur.537..374M}. To investigate this possibility, we performed high-time resolution observations to search for pulsations over a range of periods spanning milliseconds to minutes.
We observed the position of \fb\, on UTC 2019 January 30 with the ultra-wide-bandwidth (UWL) receiver deployed on the 64-m Parkes radio telescope in Australia. The output of the receiver is processed by the Medusa Graphics Processing Unit (GPU) cluster which produces 8-bit, full-Stokes filter banks with spectral and temporal resolutions of 1 MHz and 128 $\upmu$s respectively over the $3328\,\mathrm{MHz}$ band from $0.740$ - $4.03\,\mathrm{GHz}$.
The data were recorded in \textsc{psrfits} format and processed using PSRCHIVE tools \citep{HSM04}. 
Since our data were strongly affected by radio frequency interference (RFI) we used the \texttt{clfd}\footnote{\href{https://github.com/v-morello/clfd}{https://github.com/v-morello/clfd}} package described in \citep{2019MNRAS.483.3673M} to perform more sophisticated RFI mitigation.
Since radio waves are dispersed by free electrons in the interstellar medium along the line-of-sight, the data need to be corrected for this dispersion delay before any analyses can be made. 
This dispersion delay can be quantified by the dispersion measure (DM), which is the integrated free electron column density along the line-of-sight.  
We dedispersed the data over a range of trial DMs, $0.0 \leq \mathrm{DM} \leq 30.0$ pc cm$^{-3}$ in steps of 1 pc cm$^{-3}$ to account for the uncertainty in the distance to  \kstar. The DM range was estimated using various distances to the source \citep{2018AJ....156...58B}, and the NE2001 model \citep{NE2001}. 
For each trial DM, the resulting dedispersed time series was searched for long and short period pulsations using both a Fast Folding Algorithm (FFA) and a Fast Fourier Transform (FFT) respectively.

An FFT was performed using the PRESTO suite of pulsar search and analysis software \citep{REM02}. For each trial DM time series resulting from the dedispersion mentioned above, we performed an FFT using the \textsc{realfft} routine. We then used the \textsc{accelsearch} routine which sums 16 harmonics incoherently for each frequency bin to improve the signal-to-noise (S/N), and every detection above a S/N of 5 was saved for further inspection. We folded the time series for each candidate period using the \textsc{prepfold} routine and the folded time series was visually inspected to check whether it resembled a true astrophysical source. No significant periodic pulsations were detected above the threshold S/N.

While FFTs are a standard technique used for searching pulsars and short period sources \cite[e.g.][]{RCJ+18}, they become less sensitive when searching for longer periods as red noise becomes a dominant factor.
Therefore, we used the \textsc{RIPTIDE}\footnote{\href{https://bitbucket.org/vmorello/riptide/src/master/}{https://bitbucket.org/vmorello/riptide/src/master/}} FFA algorthm developed by Morello et al. (in prep) to search for periods ranging from 1 second to 10 minutes. Similar to the FFT, we dedispersed and folded each time series and vetted the candidates for significant pulse profiles above a S/N of 8. We did not detect any significant periodic pulsations above the S/N threshold.

In addition to the periodicity searches described above, we also performed offline single pulse searches on the data using the GPU based \textsc{heimdall}\footnote{\href{https://bitbucket.org/vmorello/riptide/src/master/}{https://sourceforge.net/projects/heimdall-astro/}} single pulse search software. The \textsc{heimdall} code reads in a subset of the data at a time, and processes it over the DM-width parameter space in search of single pulses. We expect any single pulse to be detected over a range of DMs close to the true value. A clustering algorithm merges these events into one before returning the DM that yields the maximum S/N. This reduces the number of candidates considerably as thousands of events can be identified prior to merging, which will then eventually result in a much lesser number of unique candidates. The data were searched for bursts in DM and width space by dedispersing over trial DMs in the range $0.0 \leq \mathrm{DM} \leq 30.0$ pc cm$^{-3}$ and convolving with a series of sliding boxcar filters (square pulse) of width $2^{0} \leq W \leq 2^{17}$ time samples respectively. No significant bursts were detected above a S/N threshold of 10.

\section{UV observations}
\label{sec: UV obs}

We were granted two target of opportunity observations of \fb\, on 2019 April 18 and 2019 May 05 with the Swift UVOT \citep{2000HEAD....5.3411M} and XRT \citep{2005SPIE.5898..325H} instruments. For the Swift UVOT observations we used the UVW1 filter (centre wavelength $2600$\,\AA). Two exposures of 393.9\,s and 1012.5\,s were taken on 2019 April 18 and three exposures of 500.25\,s, 221.68\,s and 306.54\,s were taken on 2019 May 05. The magnitudes of the source are shown in Table\,\ref{tab: UV obs}, and the UV is plotted on the SED of \kstar\, in Figure\,\ref{SED plot}.
As can be seen in Table\,\ref{tab: UV obs}, there is no evidence of variability greater than 0.02 magnitudes in the UV from these observations on timescales of a few hours and 27 days.

\begin{table}
    \caption{Details of the UV observations of \fb\, taken with the Swift UVOT instrument with the UVW1 filter (centre wavelength $2600$\,\AA).}
    \centering
    \begin{tabular}{l|l|l|l}
    \hline
     Date & Time (UTC) & Exposure (s) & AB Magnitude \\
     \hline 
     2019-04-18 & 20:45:17 & 393.93     & $15.02\pm0.02$ \\
                & 22:18:47 & 1012.48    & $15.01\pm0.02$ \\
     2019-05-05 & 03:30:48 & 500.25     & $15.00\pm0.02$ \\
                & 05:11:06 & 221.68     & $14.98\pm0.03$ \\
                & 09:41:44 & 306.54     & $15.00\pm0.02$ \\
    \hline
    \end{tabular}
    \label{tab: UV obs}
\end{table}

\section{X-ray observations}
\label{sec: X-ray obs}

The simultaneous Swift XRT \citep{2005SPIE.5898..325H} observations on the two days were 1.4\,ks and 1.0\,ks respectively.
We used the XRT photon counting mode in the 0.3--12\,keV band. During the first observing session we detected 18 events from the source. The probability that these counts are from the background is $1.1\times10^{-61}$. In the second session, 19 events were detected from the source with a probability of $1.54\times10^{-57}$ that the photons are from the background. Based on these observations we claim a faint X-ray detection of \fb. Most events detected from the source were in the soft part of the band, 0.3--3\,keV. Due to the low photon count we were unable to fit a spectrum to the individual observations, but we were able to fit a power-law to the combined spectrum from both epochs to obtain a flux density of $7.3^{+3}_{-2}\times10^{-13}\,\mathrm{erg\,cm^{-2}\,s^{-1}}$. The parameters for this fit are shown in Table~\ref{tab:spec} and one of the X-ray images is shown in Figure\,\ref{fig: radio xray uv images}.

\begin{table}
    \caption{Parameters of a power-law fit to the XRT spectrum of \fb\, with 1-sigma error bars in parentheses. $\Gamma$ is the power-law photon index, N$_{\rm H}$ is the neutral hydrogen column density along the line-of-sight. F$_{X \rm abs}$ is the absorbed X-ray flux and $\chi^{2}$ is the reduced chi-squared statistic of the fit.}
    \centering
    \begin{tabular}{l|l}
    \hline
     Parameter & Value \\
     \hline 
     $\Gamma$ &  2.4 (+0.8, -0.7)\\
     N$_{\rm H}$ (cm$^{-2}$) & 4.0(+4.0, -3.0) $\times$10$^{21}$ \\
     F$_{X\rm abs}$ (ergs~cm$^{-2}$~s$^{-1}$) &  7.3(+3, -2) $\times$10$^{-13}$\\
     $\chi^{2}$   &  2.3\\
    \hline
    \end{tabular}
    \label{tab:spec}
\end{table}

\section{Discussion}
\label{sec: discussion}

\fb\, is a transient radio source coincident with a spectroscopic binary where one of the objects is the chromospherically active K-type giant \kstar\, (spectral type K4-5). In the radio we see a bright event occur in October 2018, followed by decreasing radio brightness with some underlying variability. We do not find any radio pulsations. \fb\, was detected in both the UV and the X-ray.

The approximately 18 years of optical photometry shows that \kstar\, varies with a period of $21.25\pm0.04\,\mathrm{days}$, but that the shape and amplitude of the light curve changes over time. The optical spectra show that \fb\, is a spectroscopic binary with a period matching the photometric period. The line-of-sight radial velocity amplitude of \kstar\, is $\pm43\,\mathrm{km\,s^{-1}}$. There are no emission lines in the spectra, as would be expected for an accretion flow or accreting compact companion. The overall spectrum of the star is highly stable, but strong lines show small-scale variation consistent with an absorption component in orbital anti-phase with the K-giant. Blue lines are preferentially affected, and the combination is consistent with a fainter, broader-lined object on the other side of the barycentre, with a line-of-sight radial velocity amplitude of $12\pm10\,\mathrm{km\,s^{-1}}$ and a temperature of $\sim$7000 K. While the variations are clear, we are not able to extract this signal with sufficient clarity to state whether we have detected the signature of a spectroscopic binary companion, or some other physical effect within the system. Further spectral monitoring at high signal-to-noise will be necessary to obtain better phase coverage.

\begin{figure*}
 \includegraphics[width=\textwidth]{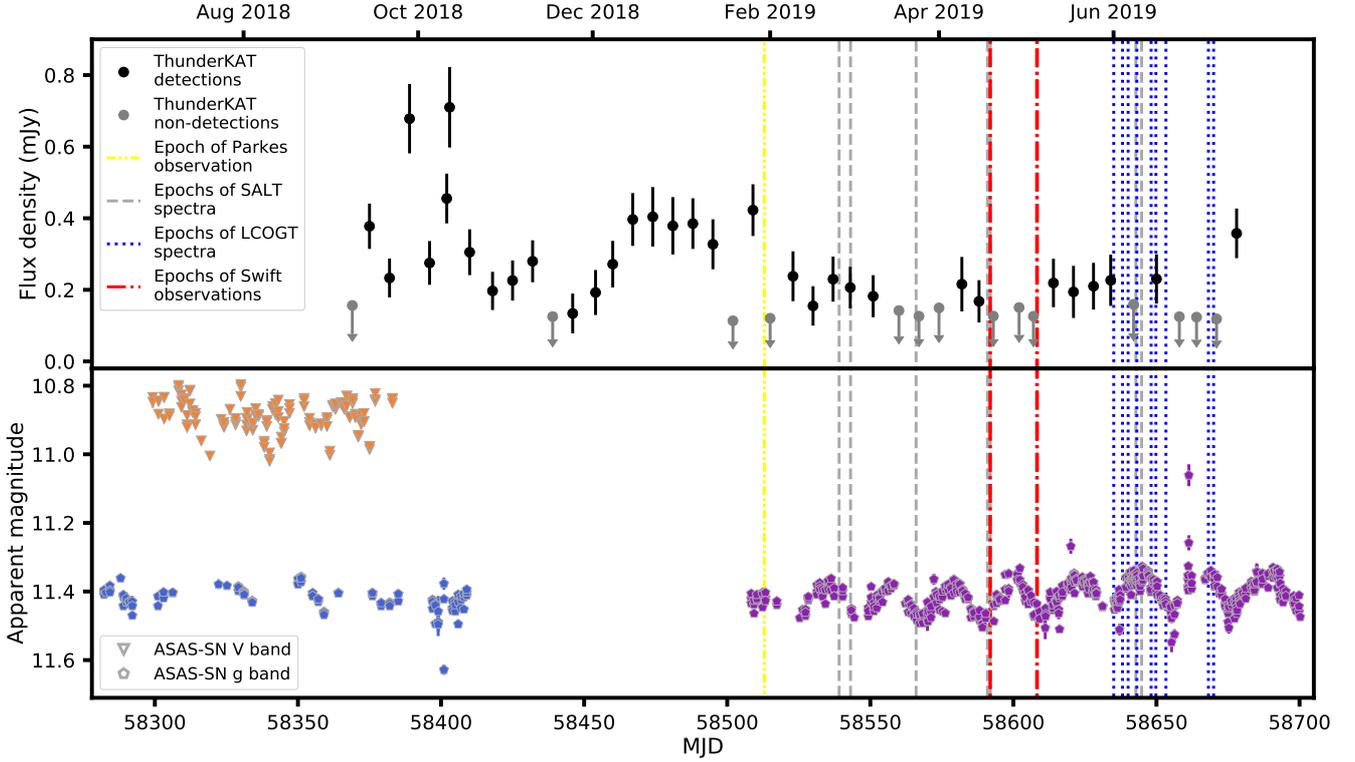}
 \caption{MeerKAT radio flux densities (top panel) and the ASAS-SN semesters (bottom panel) beginning on 2018 June 13 (MJD: 58282, g-band), 2018 June 30 (MJD: 58299, V-band) and 2019 January 25 (MJD: 58508, g-band), the colours and symbols are the same as in Figures\,\ref{fig: Optical Periods} and \ref{fig: Folded Light Curves}. The epochs of the LCO and SALT spectra, the Swift observations and the Parkes observations are plotted as vertical lines in both panels.}
 \label{fig: recent radio optical}
\end{figure*}

\begin{figure}
 \includegraphics[width=\columnwidth]{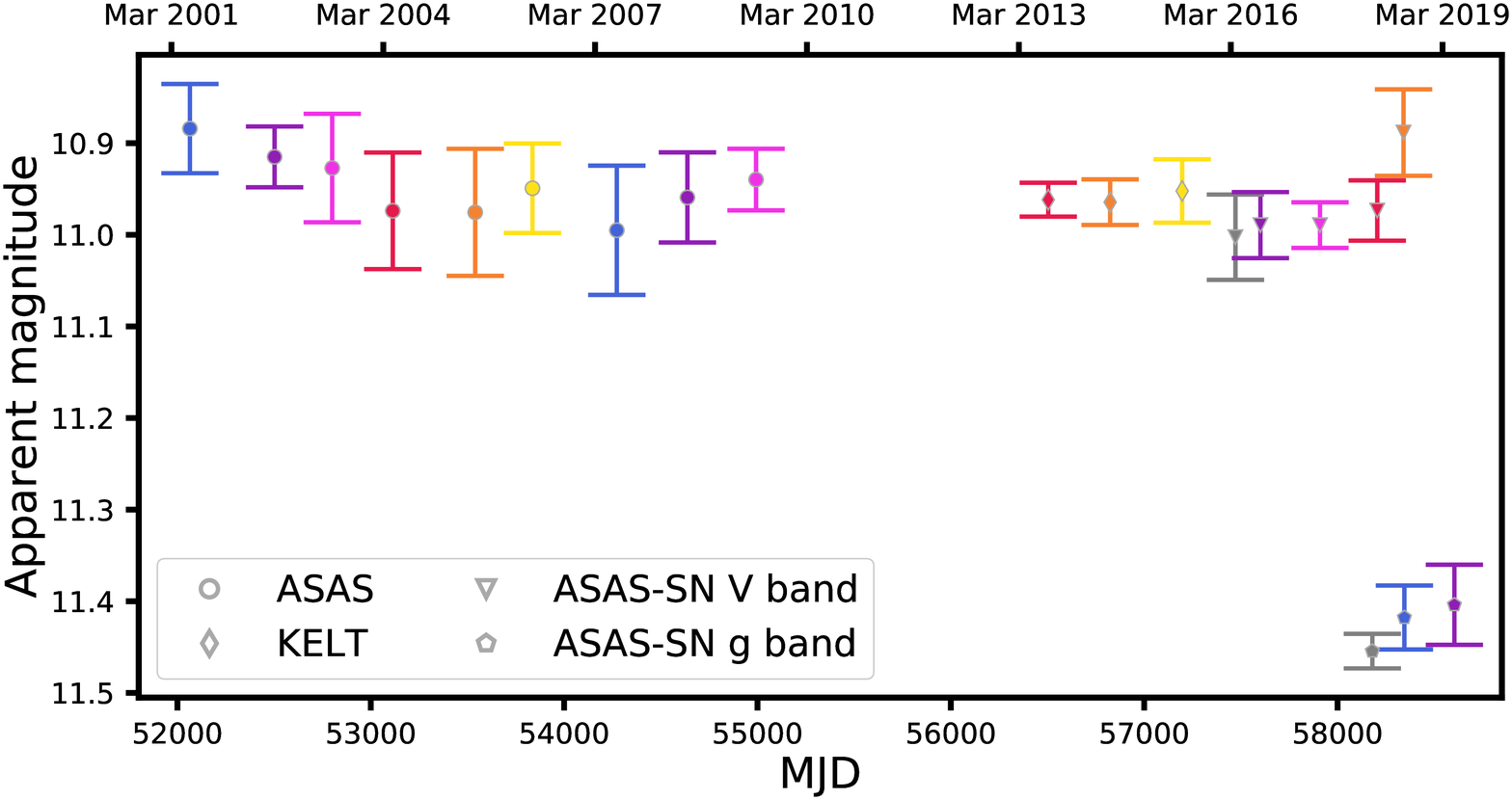}
 \caption{The weighted mean of each semester of optical observations, the bars indicate the standard deviation of the semester. The same colours and symbols are used here as in Figures\,\ref{fig: Optical Periods} and \ref{fig: Folded Light Curves}.}
 \label{fig: optical st devs}
\end{figure}

In Figure\,\ref{fig: recent radio optical} we plot the ASAS-SN semesters beginning on 2018 June 13 (MJD: 58282), 2018 June 30 (MJD: 58299) and 2019 January 25 (MJD: 58508) as well as the radio light curve from the MeerKAT observations. There is no clear periodicity in the 2018 June 30 semester of ASAS-SN V-band observations, which coincides with the bright radio event that occurred in October 2018. However, in the 2019 January 25 semester of ASAS-SN g-band observations the periodicity has returned, with  increasing amplitude over time. This is coincident with the decreasing radio flux over time. This could indicate that the cause of the optical photometric period was disrupted by the same event that caused the radio flaring in October 2018. In Figure\,\ref{fig: Folded Light Curves} we can also see that the light curve is less defined in the 2013 May 15 (MJD: 56427) and 2014 Feb 17 (MJD: 56705) semester, but that the periodicity returns in the 2015 Mar 03 (MJD: 57084) semester. This could imply that a similar outburst to the October 2018 event occurred before or during the 2013 May 15 semester. A search of archival radio data has not revealed any data at these times to see whether there was a corresponding radio outburst. We also see, in Figure\,\ref{fig: optical st devs}, that the optical brightness of the star and the amplitude of the variability varies over time. Particularly that the 2018 June 30 semester of ASAS-SN V-band observations, that ended in October 2018 during the radio flare, are significantly brighter than previous semesters of optical photometry. We also note that there appears to be a flare in the ASAS-SN g-band observations in June 2019; however, this is an instrumental issue as the stars nearby to \kstar\, show a similar ``flare'' in the same epoch in their light curves.

\begin{figure}
 \includegraphics[width=\columnwidth]{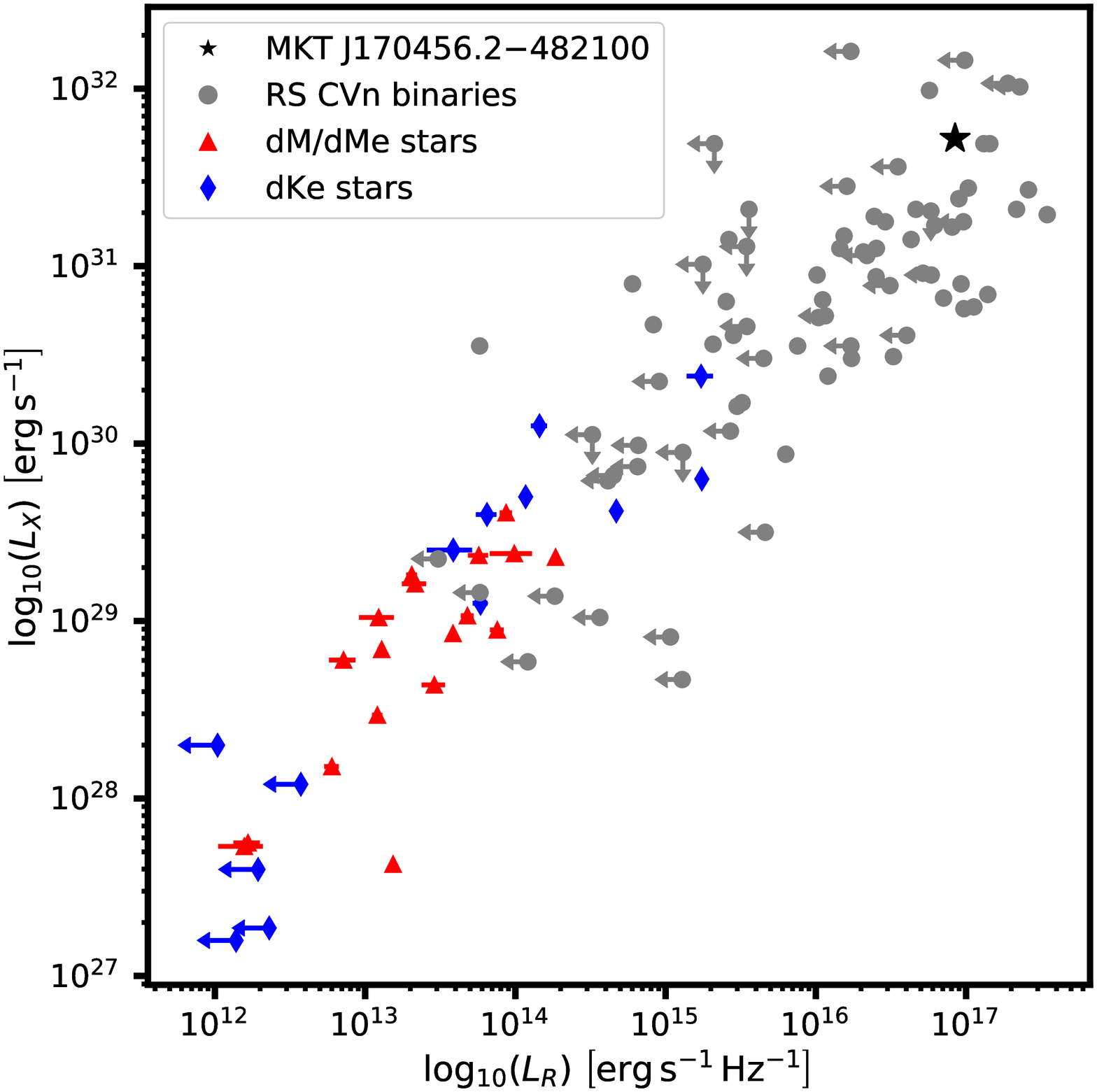}
 \caption{Plot of the average soft X-ray and radio luminosity of active stars from \citet{1993ApJ...405L..63G,1994A&A...285..621B} and references therein. dM, dMe, and dKe stars are magnetically active stars. The uncertainties were not available on many of these archival data points, but are included where possible. The X-ray luminosity for \fb\, was found by fitting a spectrum to all of the Swift XRT observations we have obtained, for the Swift 0.3--12\,keV band. The \fb\, radio luminosity is the weighted mean of the radio luminosity across all epochs. There are uncertainties on the \fb\, values, but they are too small to be seen. }
 \label{fig: BG plot}
\end{figure}

The variable shape of the photometric light curve could indicate the presence of large star spots on the surface of the K star \citep[e.g.][]{2005A&A...432..647F,2006A&A...446.1129B}. For example, the shape of the V-band light curve in the latest semester of ASAS-SN g-band observations could be caused by the presence of two large star spots, similar to IM Peg \citep{2006A&A...446.1129B}. The variability in the shape and phase of the light curve over time could indicate that the spots change in size, number, and position on the star's surface. 
Assuming that the variability is caused by star spots, we can calculate the covering fraction, i.e. the fraction of the star covered by star spots using \citep{2018AJ....156..203M}:
\begin{equation}
    f_{S,\mathrm{min}} = \frac{1-\mathrm{min}\mathcal{F}}{1-c}
\end{equation}
where $\mathcal{F}$ is the flux of the star divided by the mean flux and $c$ is the spot contrast. A spot with the same intensity as the photosphere of the star would have $c=0$, while a completely dark spot would have $c=1$ \citep{2018AJ....156..203M}. Sunspots have a contrast value of $c=0.3$. The covering fractions for the ASAS-SN V- and g-band semesters are shown in Table\,\ref{tab: spot filling}, suggesting that $\gtrsim$10 per cent of the stellar disc is spotted as a temporal average. If \kstar\, is spotted, the radio activity and X-ray observations could also be from the star, similar to an RS CVn system.
\citet{1993ApJ...405L..63G} showed that there is a correlation between the average soft X-ray luminosity and radio luminosity for active stars. If we place the average radio luminosity and X-ray luminosity of \fb\, on the Benz-G\"udel plot in Figure\,\ref{fig: BG plot}, then it fits well with the known values for RS CVn systems.

\begin{table}
    \caption{The covering fraction, $f_{S,\min}$, of spots on the K-star assuming different values for $c$, the spot contrast. These values were calculated using the ASAS-SN V- and g-band observations. The semester start date is the date of the first observation of that semester.}
    \centering
    \begin{tabular}{l|p{1.5cm}|l|l|l|l}
    \hline
    Band & Semester start date & \multicolumn{4}{c}{$f_{S,min}$} \\
         &                     & $c=0.0$ & $c=0.3$ & $c=0.5$ & $c=0.9$ \\
    \hline 
    V  & 2016-03-10 & 0.12 & 0.17 & 0.24 & 1.21 \\
    V  & 2016-05-21 & 0.07 & 0.09 & 0.13 & 0.66 \\
    V  & 2017-01-24 & 0.18 & 0.27 & 0.37 & 1.87 \\
    V  & 2018-02-01 & 0.14 & 0.20 & 0.27 & 1.37 \\
    g  & 2018-02-26 & 0.05 & 0.08 & 0.11 & 0.54 \\
    g  & 2018-06-13 & 0.17 & 0.24 & 0.34 & 1.70 \\
    V  & 2018-06-30 & 0.11 & 0.16 & 0.23 & 1.14 \\
    g  & 2019-01-25 & 0.15 & 0.22 & 0.31 & 1.55 \\
    \hline
    \end{tabular}
    \label{tab: spot filling}
\end{table}

The minimum mass for a star to have evolved into a sub-giant in $13.8\,\mathrm{Gyr}$ is $0.935\,\mathrm{M_{\odot}}$ \citep[e.g.][]{2017ApJ...835...77M,2014MNRAS.445.4287T,2015MNRAS.452.1068C,2014MNRAS.444.2525C}. \kstar\, is not significantly metal-poor, but to take into account possible metallicity effects \citep{2015MNRAS.448..502M} we use a conservative minimum mass for the star of $0.8\,\mathrm{M_{\odot}}$. We know from the spectra that the velocity amplitude of \kstar\, is $43\,\mathrm{km\,s^{-1}}$, and if we assume, from the sinusoidal shape of the radial velocity curve shown in Figure\,\ref{rvcurve}, that the orbit is circular we can then use the mass function to calculate the mass of the companion ($m_2$):
\begin{equation}
    \frac{m_{2}^{3}}{\left(m_1 + m_2\right)^2}\sin{i}^3 = \frac{P}{2\pi G}v_{1}^{3}
\end{equation}
where $m_1$ and $v_1$ are the mass and radial velocity amplitude of \kstar\, respectively, $i$ is the inclination of the binary system, and $P$ is the orbital period.
Assuming an inclination of $90^{\circ}$, a period of $21.25\pm0.04\,\mathrm{days}$ from the optical photometry, and a K-star mass of $0.8\,\mathrm{M_{\odot}}$  gives a minimum companion mass of $\sim0.75\,\mathrm{M_{\odot}}$ and a semi-major axis of $\sim0.17\,\mathrm{AU}$. If the mass of \kstar\, is between $\sim2.22$ and $\sim2.48\,\mathrm{M_{\odot}}$ (Section \ref{SEDsect}), then an inclination of $90^{\circ}$ gives a minimum companion mass of $1.29\,\mathrm{M_{\odot}}$. These values show that the companion may be a white dwarf.

However, if we interpret the spectral variations we see as absorption lines from a companion star on the opposite side of the barycentre to \kstar, its low radial velocity amplitude ($\sim12\pm10\,\mathrm{km\,s^{-1}}$; see Sec.\,\ref{sec: spectral var}) requires some explanation. 
The binary companion to \kstar\, does not significantly affect the shape of the SED (Section \ref{SEDsect}; Figure \ref{SED plot}). The blue and UV photometry in particular mean that it is intrinsically much fainter and/or has a temperature less than $\sim7000\,\mathrm{K}$. If the companion is a main-sequence star, this combination restricts its mass to $\lesssim1.5\,\mathrm{M_{\odot}}$.

Still assuming that the $\sim12\pm10\,\mathrm{km\,s^{-1}}$ is from a companion to \kstar, we can take the ratio of the velocities to find a mass ratio. 
We then find that the mass of the companion is between $\sim2 \times M_{\mathrm{K\,giant}}$ and $\sim22 \times M_{\mathrm{K\,giant}}$. 
Taking our conservative minimum mass of \kstar, $\sim0.8\,\mathrm{M_{\odot}}$, provides a companion mass between $\sim1.6$ and $\sim17.2\,\mathrm{M_{\odot}}$, which corresponds to a semi-major axis, for a circular orbit, of $0.20\,\mathrm{AU}$ and $0.40\,\mathrm{AU}$ respectively. Our modelled mass of \kstar\, $2.22\sim2.48\,\mathrm{M_{\odot}}$, 
would mean a companion mass between $\sim4.3$ and $\sim53.3\,\mathrm{M_{\odot}}$, which corresponds to a semi-major axis of $0.28\,\mathrm{AU}$ and $0.57\,\mathrm{AU}$ respectively. These masses are not compatible with a white dwarf, or with a star with a temperature less than $\sim7000\,\mathrm{K}$, and would lead to significant tidal distortion of \kstar, which would be visible in the optical lightcurve.


One possible explanation is that \fb\, is a triple-system, where the companion to \kstar\, is actually two objects orbiting each other, with the semi-major axis ($a_{\mathrm{inner}}$) of the inner orbit much smaller than the semi-major axis ($a_{\mathrm{outer}}$) of the outer orbit. If we assume that the inner orbit is circular with an inclination of $90^{\circ}$, and is comprised of two objects of equal mass (and hence equal radial velocity), we can calculate the minimum radial velocity amplitude of those objects. If we assume the minimum masses from above ($M_{\mathrm{K\,giant}}=0.8\,\mathrm{M_{\odot}}$, $M_{\mathrm{companion}}=1.6\,\mathrm{M_{\odot}}$, $a_{\mathrm{outer}}=0.2\,\mathrm{AU}$), and a semi-major axis of $<10$ per cent of the outer orbit ($a_{\mathrm{inner}}=0.02\,\mathrm{AU}$), then the velocity amplitude of one object in the inner orbit is $\gtrsim130\,\mathrm{km\,s^{-1}}$. This is inconsistent with a radial velocity amplitude of $12\pm10\,\mathrm{km\,s^{-1}}$, and the spectral variations we see would reflect these higher velocity amplitudes.

Further investigation into this system is required to determine the nature of the companion to \kstar, as well as to confirm the origin of the X-ray, radio, and UV flux. As \kstar\, is chromospherically active, the X-ray, radio and UV emission may be from this star. A UV spectrum would confirm the nature of the UV emission. This MeerKAT field is being observed weekly by ThunderKAT until September 2023, and ASAS-SN continues to observe this field. This will help to determine whether the radio and optical variability is related, which will help determine whether the radio activity is from flaring activity on \kstar. We are also observing \fb\, weekly in the X-ray for 9 weeks with Swift. This will help determine whether the source is variable in the X-ray. \emph{TESS} observations of the source will be released later in 2019, which may reveal shorter timescale optical variability.

\section{Conclusions}
\label{sec: conclusion}

We report the discovery of \fb, the first transient discovered with MeerKAT. \fb\, coincides with the K4-5V type sub-giant star \kstar. 
Using $\sim18\,\mathrm{years}$ of optical photometry from ASAS, KELT, and ASAS-SN we find that \kstar\, has a photometric period of $21.25\pm0.04\,\mathrm{days}$. 
A model of the light curve which attributes the variability to the presence of stars spots explains the light curve shape, changing phase, and results in a reasonable covering fraction. 
Using spectra from SALT and LCO we find that the star is in a spectroscopic binary, with a velocity amplitude of $43\,\mathrm{km\,s^{-1}}$. 
The X-ray and UV flux that we detect from the position of \fb\, indicates that this flux and the radio variability may be because the system is an RS CVn type system.
There are absorption lines indicating spectral variation within the system, in anti-phase to the K-giant spectrum, at a radial velocity amplitude of $12\pm10\,\mathrm{km\,s^{-1}}$. However, we lack sufficient data to identify this with a causative mechanism, as binary companions at this radial velocity do not make physical sense.

\section*{Acknowledgements}

LND, MC, KMR, and BWS acknowledge support from the European Research Council (ERC) under the European Union's Horizon 2020 research and innovation programme (grant agreement No 694745). 
This work was partly funded by the United Kingdom's Science and Technologies Facilities Council (STFC) grant ST/P000649/1.
DAHB acknowledges research support from the South African National Research Foundation. 
ET acknowledges financial support from the UnivEarthS Labex program of Sorbonne Paris Cit\'{e} (ANR-10-LABX-0023 and ANR-11-IDEX-0005-02).
PAW acknowledges support from the National Research Foundation (NRF) and the University of Cape Town (UCT).
This research was supported by a Grant from the GIF, the German-Israeli Foundation for Scientific Research and Development. A.H. Acknowledges support from GIF.
JCAM-J is the recipient of an Australian Research Council Future Fellowship (FT140404082), funded by the Australian Government.
We acknowledge use of the Inter-University Institute for Data Intensive Astronomy (IDIA) data intensive research cloud for data processing. IDIA is a South African university partnership involving the University of Cape Town, the University of Pretoria and the University of the Western Cape.
The MeerKAT telescope is operated by the South African Radio Astronomy Observatory (SARAO), which is a facility of the National Research Foundation, an agency of the Department of Science and Technology. We would like to thank the operators, SARAO staff and ThunderKAT Large Survey Project team. The Parkes radio telescope is part of the Australia Telescope National Facility which is funded by the Commonwealth of Australia for operation as a National Facility managed by CSIRO. Some of these observations were obtained with the Southern African Large Telescope under the Large Science Programme on transients, 2018-2-LSP-001 (PI: DAHB). 
This work makes use of observations from the LCO network.
KELT data are made available to the community through the Exoplanet Archive on behalf of the KELT project team, kindly provided by Rudi Kuhn.
This work has made use of data from the European Space Agency (ESA) mission
{\it Gaia}.\footnote{\href{https://www.cosmos.esa.int/gaia}{https://www.cosmos.esa.int/gaia}, processed by the {\it Gaia}
Data Processing and Analysis Consortium (DPAC,
\href{https://www.cosmos.esa.int/web/gaia/dpac/consortium}{https://www.cosmos.esa.int/web/gaia/dpac/consortium})} Funding for the DPAC
has been provided by national institutions, in particular the institutions
participating in the {\it Gaia} Multilateral Agreement.
This research made use of Astropy,\footnote{http://www.astropy.org} a community-developed core Python package for Astronomy \citep{astropy:2013, astropy:2018}.
This research made use of APLpy, an open-source plotting package for Python \citep{2012ascl.soft08017R}.
LND would like to thank Colin Clark, Mark Kennedy and Daniel Mata for useful and interesting discussions.
We would like to thank the referee for their constructive comments on the manuscript.




\bibliographystyle{mnras}
\bibliography{TYC} 

\begin{thebibliography}{}
\makeatletter
\relax
\def\mn@urlcharsother{\let\do\@makeother \do\$\do\&\do\#\do\^\do\_\do\%\do\~}
\def\mn@doi{\begingroup\mn@urlcharsother \@ifnextchar [ {\mn@doi@}
  {\mn@doi@[]}}
\def\mn@doi@[#1]#2{\def\@tempa{#1}\ifx\@tempa\@empty \href
  {http://dx.doi.org/#2} {doi:#2}\else \href {http://dx.doi.org/#2} {#1}\fi
  \endgroup}
\def\mn@eprint#1#2{\mn@eprint@#1:#2::\@nil}
\def\mn@eprint@arXiv#1{\href {http://arxiv.org/abs/#1} {{\tt arXiv:#1}}}
\def\mn@eprint@dblp#1{\href {http://dblp.uni-trier.de/rec/bibtex/#1.xml}
  {dblp:#1}}
\def\mn@eprint@#1:#2:#3:#4\@nil{\def\@tempa {#1}\def\@tempb {#2}\def\@tempc
  {#3}\ifx \@tempc \@empty \let \@tempc \@tempb \let \@tempb \@tempa \fi \ifx
  \@tempb \@empty \def\@tempb {arXiv}\fi \@ifundefined
  {mn@eprint@\@tempb}{\@tempb:\@tempc}{\expandafter \expandafter \csname
  mn@eprint@\@tempb\endcsname \expandafter{\@tempc}}}

\bibitem[\protect\citeauthoryear{{Allard}, {Guillot}, {Ludwig}, {Hauschildt},
  {Schweitzer}, {Alexander}  \& {Ferguson}}{{Allard}
  et~al.}{2003}]{2003IAUS..211..325A}
{Allard} F.,  {Guillot} T.,  {Ludwig} H.-G.,  {Hauschildt} P.~H.,  {Schweitzer}
  A.,  {Alexander} D.~R.,   {Ferguson} J.~W.,  2003, in {E.~Mart{\'{\i}}n} ed.,
   IAU Symposium Vol. 211, Brown Dwarfs. p.~325

\bibitem[\protect\citeauthoryear{{Astropy Collaboration} et~al.,}{{Astropy
  Collaboration} et~al.}{2013}]{astropy:2013}
{Astropy Collaboration} et~al., 2013, \mn@doi [\aap]
  {10.1051/0004-6361/201322068}, \href
  {http://adsabs.harvard.edu/abs/2013A%26A...558A..33A} {558, A33}

\bibitem[\protect\citeauthoryear{{Bailer-Jones}, {Rybizki}, {Fouesneau},
  {Mantelet}  \& {Andrae}}{{Bailer-Jones} et~al.}{2018}]{2018AJ....156...58B}
{Bailer-Jones} C.~A.~L.,  {Rybizki} J.,  {Fouesneau} M.,  {Mantelet} G.,
  {Andrae} R.,  2018, \mn@doi [\aj] {10.3847/1538-3881/aacb21}, \href
  {http://adsabs.harvard.edu/abs/2018AJ....156...58B} {156, 58}

\bibitem[\protect\citeauthoryear{{Bannister}, {Murphy}, {Gaensler}, {Hunstead}
  \& {Chatterjee}}{{Bannister} et~al.}{2011}]{2011MNRAS.412..634B}
{Bannister} K.~W.,  {Murphy} T.,  {Gaensler} B.~M.,  {Hunstead} R.~W.,
  {Chatterjee} S.,  2011, \mn@doi [\mnras] {10.1111/j.1365-2966.2010.17938.x},
  \href {https://ui.adsabs.harvard.edu/abs/2011MNRAS.412..634B} {412, 634}

\bibitem[\protect\citeauthoryear{{Benz} \& {Guedel}}{{Benz} \&
  {Guedel}}{1994}]{1994A&A...285..621B}
{Benz} A.~O.,  {Guedel} M.,  1994, \aap, \href
  {https://ui.adsabs.harvard.edu/abs/1994A&A...285..621B} {285, 621}

\bibitem[\protect\citeauthoryear{{Bhandari} et~al.,}{{Bhandari}
  et~al.}{2018}]{2018MNRAS.478.1784B}
{Bhandari} S.,  et~al., 2018, \mn@doi [\mnras] {10.1093/mnras/sty1157}, \href
  {https://ui.adsabs.harvard.edu/abs/2018MNRAS.478.1784B} {478, 1784}

\bibitem[\protect\citeauthoryear{{Biazzo}, {Frasca}, {Catalano}  \&
  {Marilli}}{{Biazzo} et~al.}{2006}]{2006A&A...446.1129B}
{Biazzo} K.,  {Frasca} A.,  {Catalano} S.,   {Marilli} E.,  2006, \mn@doi
  [\aap] {10.1051/0004-6361:20053213}, \href
  {https://ui.adsabs.harvard.edu/abs/2006A&A...446.1129B} {446, 1129}

\bibitem[\protect\citeauthoryear{{Bower}, {Saul}, {Bloom}, {Bolatto},
  {Filippenko}, {Foley}  \& {Perley}}{{Bower}
  et~al.}{2007}]{2007ApJ...666..346B}
{Bower} G.~C.,  {Saul} D.,  {Bloom} J.~S.,  {Bolatto} A.,  {Filippenko} A.~V.,
  {Foley} R.~J.,   {Perley} D.,  2007, \mn@doi [\apj] {10.1086/519831}, \href
  {https://ui.adsabs.harvard.edu/abs/2007ApJ...666..346B} {666, 346}

\bibitem[\protect\citeauthoryear{{Bramall} et~al.,}{{Bramall}
  et~al.}{2012}]{Bramall2012}
{Bramall} D.~G.,  et~al., 2012, in Ground-based and Airborne Instrumentation
  for Astronomy IV. p. 84460A, \mn@doi{10.1117/12.925935}

\bibitem[\protect\citeauthoryear{{Bright} et~al.,}{{Bright}
  et~al.}{2019}]{2019MNRAS.486.2721B}
{Bright} J.~S.,  et~al., 2019, \mn@doi [\mnras] {10.1093/mnras/stz1004}, \href
  {https://ui.adsabs.harvard.edu/abs/2019MNRAS.486.2721B} {486, 2721}

\bibitem[\protect\citeauthoryear{{Brown} et~al.,}{{Brown}
  et~al.}{2013}]{2013PASP..125.1031B}
{Brown} T.~M.,  et~al., 2013, \mn@doi [\pasp] {10.1086/673168}, \href
  {https://ui.adsabs.harvard.edu/abs/2013PASP..125.1031B} {125, 1031}

\bibitem[\protect\citeauthoryear{{Buckley}, {Swart}  \& {Meiring}}{{Buckley}
  et~al.}{2006}]{Buckley2006}
{Buckley} D. A.~H.,  {Swart} G.~P.,   {Meiring} J.~G.,  2006, in Society of
  Photo-Optical Instrumentation Engineers (SPIE) Conference Series. p. 62670Z,
  \mn@doi{10.1117/12.673750}

\bibitem[\protect\citeauthoryear{{Buckley}, {Meintjes}, {Potter}, {Marsh}  \&
  {G{\"a}nsicke}}{{Buckley} et~al.}{2017}]{2017NatAs...1E..29B}
{Buckley} D.~A.~H.,  {Meintjes} P.~J.,  {Potter} S.~B.,  {Marsh} T.~R.,
  {G{\"a}nsicke} B.~T.,  2017, \mn@doi [Nature Astronomy]
  {10.1038/s41550-016-0029}, \href
  {https://ui.adsabs.harvard.edu/abs/2017NatAs...1E..29B} {1, 0029}

\bibitem[\protect\citeauthoryear{{Burgh}, {Nordsieck}, {Kobulnicky},
  {Williams}, {O'Donoghue}, {Smith}  \& {Percival}}{{Burgh}
  et~al.}{2003}]{Burgh2003}
{Burgh} E.~B.,  {Nordsieck} K.~H.,  {Kobulnicky} H.~A.,  {Williams} T.~B.,
  {O'Donoghue} D.,  {Smith} M.~P.,   {Percival} J.~W.,  2003, in {Iye} M.,
  {Moorwood} A. F.~M.,  eds,  Society of Photo-Optical Instrumentation
  Engineers (SPIE) Conference Series Vol. 4841, Instrument Design and
  Performance for Optical/Infrared Ground-based Telescopes. pp 1463--1471,
  \mn@doi{10.1117/12.460312}

\bibitem[\protect\citeauthoryear{{Camilo} et~al.,}{{Camilo}
  et~al.}{2018}]{2018ApJ...856..180C}
{Camilo} F.,  et~al., 2018, \mn@doi [\apj] {10.3847/1538-4357/aab35a}, \href
  {http://adsabs.harvard.edu/abs/2018ApJ...856..180C} {856, 180}

\bibitem[\protect\citeauthoryear{{Carbone} et~al.,}{{Carbone}
  et~al.}{2016}]{2016MNRAS.459.3161C}
{Carbone} D.,  et~al., 2016, \mn@doi [\mnras] {10.1093/mnras/stw539}, \href
  {https://ui.adsabs.harvard.edu/abs/2016MNRAS.459.3161C} {459, 3161}

\bibitem[\protect\citeauthoryear{{Chandra} \& {Frail}}{{Chandra} \&
  {Frail}}{2012}]{2012ApJ...746..156C}
{Chandra} P.,  {Frail} D.~A.,  2012, \mn@doi [\apj]
  {10.1088/0004-637X/746/2/156}, \href
  {https://ui.adsabs.harvard.edu/abs/2012ApJ...746..156C} {746, 156}

\bibitem[\protect\citeauthoryear{{Chen}, {Girardi}, {Bressan}, {Marigo},
  {Barbieri}  \& {Kong}}{{Chen} et~al.}{2014}]{2014MNRAS.444.2525C}
{Chen} Y.,  {Girardi} L.,  {Bressan} A.,  {Marigo} P.,  {Barbieri} M.,   {Kong}
  X.,  2014, \mn@doi [\mnras] {10.1093/mnras/stu1605}, \href
  {https://ui.adsabs.harvard.edu/abs/2014MNRAS.444.2525C} {444, 2525}

\bibitem[\protect\citeauthoryear{{Chen}, {Bressan}, {Girardi}, {Marigo}, {Kong}
   \& {Lanza}}{{Chen} et~al.}{2015}]{2015MNRAS.452.1068C}
{Chen} Y.,  {Bressan} A.,  {Girardi} L.,  {Marigo} P.,  {Kong} X.,   {Lanza}
  A.,  2015, \mn@doi [\mnras] {10.1093/mnras/stv1281}, \href
  {https://ui.adsabs.harvard.edu/abs/2015MNRAS.452.1068C} {452, 1068}

\bibitem[\protect\citeauthoryear{{Coppejans} et~al.,}{{Coppejans}
  et~al.}{2016}]{2016MNRAS.463.2229C}
{Coppejans} D.~L.,  et~al., 2016, \mn@doi [\mnras] {10.1093/mnras/stw2133},
  \href {https://ui.adsabs.harvard.edu/abs/2016MNRAS.463.2229C} {463, 2229}

\bibitem[\protect\citeauthoryear{{Cordes} \& {Lazio}}{{Cordes} \&
  {Lazio}}{2002}]{NE2001}
{Cordes} J.~M.,  {Lazio} T.~J.~W.,  2002, arXiv e-prints, \href
  {https://ui.adsabs.harvard.edu/abs/2002astro.ph..7156C} {pp
  astro--ph/0207156}

\bibitem[\protect\citeauthoryear{{Craig} et~al.,}{{Craig}
  et~al.}{1997}]{1997ApJS..113..131C}
{Craig} N.,  et~al., 1997, \mn@doi [\apjs] {10.1086/313052}, \href
  {https://ui.adsabs.harvard.edu/abs/1997ApJS..113..131C} {113, 131}

\bibitem[\protect\citeauthoryear{{Crause} et~al.,}{{Crause}
  et~al.}{2014}]{Crause2014}
{Crause} L.~A.,  et~al., 2014, in Ground-based and Airborne Instrumentation for
  Astronomy V. p. 91476T, \mn@doi{10.1117/12.2055635}

\bibitem[\protect\citeauthoryear{{Crawford} et~al.,}{{Crawford}
  et~al.}{2016}]{Crawford2016}
{Crawford} S.~M.,  et~al., 2016, in Ground-based and Airborne Instrumentation
  for Astronomy VI. p. 99082L, \mn@doi{10.1117/12.2232653}

\bibitem[\protect\citeauthoryear{{Currie}, {Berry}, {Jenness}, {Gibb}, {Bell}
  \& {Draper}}{{Currie} et~al.}{2014}]{2014ASPC..485..391C}
{Currie} M.~J.,  {Berry} D.~S.,  {Jenness} T.,  {Gibb} A.~G.,  {Bell} G.~S.,
  {Draper} P.~W.,  2014, in {Manset} N.,  {Forshay} P.,  eds,  Astronomical
  Society of the Pacific Conference Series Vol. 485, Astronomical Data Analysis
  Software and Systems XXIII. p.~391

\bibitem[\protect\citeauthoryear{{Cutri} et~al.,}{{Cutri}
  et~al.}{2003}]{2003tmc..book.....C}
{Cutri} R.~M.,  et~al., 2003, {2MASS All Sky Catalog of point sources.}.
The IRSA 2MASS All-Sky Point Source Catalog, NASA/IPAC Infrared Science
  Archive.

\bibitem[\protect\citeauthoryear{{Cutri} et~al.,}{{Cutri}
  et~al.}{2013}]{2013wise.rept....1C}
{Cutri} R.~M.,  et~al., 2013, Technical report, {Explanatory Supplement to the
  AllWISE Data Release Products}

\bibitem[\protect\citeauthoryear{{Draine}}{{Draine}}{2003}]{2003ApJ...598.1017D}
{Draine} B.~T.,  2003, \mn@doi [\apj] {10.1086/379118}, \href
  {http://adsabs.harvard.edu/abs/2003ApJ...598.1017D} {598, 1017}

\bibitem[\protect\citeauthoryear{{Fender} et~al.,}{{Fender}
  et~al.}{2017}]{2017arXiv171104132F}
{Fender} R.,  et~al., 2017, arXiv e-prints, \href
  {https://ui.adsabs.harvard.edu/abs/2017arXiv171104132F} {p. arXiv:1711.04132}

\bibitem[\protect\citeauthoryear{{Fong}, {Berger}, {Margutti}  \&
  {Zauderer}}{{Fong} et~al.}{2015}]{2015ApJ...815..102F}
{Fong} W.,  {Berger} E.,  {Margutti} R.,   {Zauderer} B.~A.,  2015, \mn@doi
  [\apj] {10.1088/0004-637X/815/2/102}, \href
  {https://ui.adsabs.harvard.edu/abs/2015ApJ...815..102F} {815, 102}

\bibitem[\protect\citeauthoryear{{Fouqu{\'e}} et~al.,}{{Fouqu{\'e}}
  et~al.}{2000}]{2000A&AS..141..313F}
{Fouqu{\'e}} P.,  et~al., 2000, \mn@doi [Astronomy and Astrophysics Supplement
  Series] {10.1051/aas:2000123}, \href
  {https://ui.adsabs.harvard.edu/abs/2000A&AS..141..313F} {141, 313}

\bibitem[\protect\citeauthoryear{{Frail}, {Kulkarni}, {Ofek}, {Bower}  \&
  {Nakar}}{{Frail} et~al.}{2012}]{2012ApJ...747...70F}
{Frail} D.~A.,  {Kulkarni} S.~R.,  {Ofek} E.~O.,  {Bower} G.~C.,   {Nakar} E.,
  2012, \mn@doi [\apj] {10.1088/0004-637X/747/1/70}, \href
  {https://ui.adsabs.harvard.edu/abs/2012ApJ...747...70F} {747, 70}

\bibitem[\protect\citeauthoryear{{Frasca}, {Biazzo}, {Catalano}, {Marilli},
  {Messina}  \& {Rodon{\`o}}}{{Frasca} et~al.}{2005}]{2005A&A...432..647F}
{Frasca} A.,  {Biazzo} K.,  {Catalano} S.,  {Marilli} E.,  {Messina} S.,
  {Rodon{\`o}} M.,  2005, \mn@doi [\aap] {10.1051/0004-6361:20041373}, \href
  {https://ui.adsabs.harvard.edu/abs/2005A&A...432..647F} {432, 647}

\bibitem[\protect\citeauthoryear{{Gaia Collaboration} et~al.,}{{Gaia
  Collaboration} et~al.}{2016}]{2016A&A...595A...1G}
{Gaia Collaboration} et~al., 2016, \mn@doi [\aap]
  {10.1051/0004-6361/201629272}, \href
  {https://ui.adsabs.harvard.edu/abs/2016A%26A...595A...1G} {595, A1}

\bibitem[\protect\citeauthoryear{{Gaia Collaboration} et~al.,}{{Gaia
  Collaboration} et~al.}{2018}]{2018A&A...616A...1G}
{Gaia Collaboration} et~al., 2018, \mn@doi [\aap]
  {10.1051/0004-6361/201833051}, \href
  {https://ui.adsabs.harvard.edu/abs/2018A%26A...616A...1G} {616, A1}

\bibitem[\protect\citeauthoryear{{Garc{\'\i}a-S{\'a}nchez}, {Paredes}  \&
  {Rib{\'o}}}{{Garc{\'\i}a-S{\'a}nchez} et~al.}{2003}]{2003A&A...403..613G}
{Garc{\'\i}a-S{\'a}nchez} J.,  {Paredes} J.~M.,   {Rib{\'o}} M.,  2003, \mn@doi
  [\aap] {10.1051/0004-6361:20030361}, \href
  {https://ui.adsabs.harvard.edu/abs/2003A&A...403..613G} {403, 613}

\bibitem[\protect\citeauthoryear{{Guedel} \& {Benz}}{{Guedel} \&
  {Benz}}{1993}]{1993ApJ...405L..63G}
{Guedel} M.,  {Benz} A.~O.,  1993, \mn@doi [\apjl] {10.1086/186766}, \href
  {https://ui.adsabs.harvard.edu/abs/1993ApJ...405L..63G} {405, L63}

\bibitem[\protect\citeauthoryear{{Gunn}}{{Gunn}}{1996}]{1996IrAJ...23..137G}
{Gunn} A.~G.,  1996, Irish Astronomical Journal, \href
  {https://ui.adsabs.harvard.edu/abs/1996IrAJ...23..137G} {23, 137}

\bibitem[\protect\citeauthoryear{{Hall}}{{Hall}}{1976}]{1976ASSL...60..287H}
{Hall} D.~S.,  1976, in {Fitch} W.~S.,  ed.,  Astrophysics and Space Science
  Library Vol. 60, IAU Colloq. 29: Multiple Periodic Variable Stars. p.~287,
  \mn@doi{10.1007/978-94-010-1175-4_15}

\bibitem[\protect\citeauthoryear{Hallinan, Mooley, Kulkarniet  et~al.}{Hallinan
  et~al.}{2013}]{hallinan2013transient}
Hallinan G.,  Mooley K.,  Kulkarniet S.,   et~al., 2013, VLASS White Paper

\bibitem[\protect\citeauthoryear{{Hallinan} et~al.,}{{Hallinan}
  et~al.}{2017}]{2017Sci...358.1579H}
{Hallinan} G.,  et~al., 2017, \mn@doi [Science] {10.1126/science.aap9855},
  \href {http://adsabs.harvard.edu/abs/2017Sci...358.1579H} {358, 1579}

\bibitem[\protect\citeauthoryear{{Hernandez}, {Zharikov}, {Neustroev}  \&
  {Tovmassian}}{{Hernandez} et~al.}{2017}]{2017MNRAS.470.1960H}
{Hernandez} M.~S.,  {Zharikov} S.,  {Neustroev} V.,   {Tovmassian} G.,  2017,
  \mn@doi [\mnras] {10.1093/mnras/stx1341}, \href
  {https://ui.adsabs.harvard.edu/abs/2017MNRAS.470.1960H} {470, 1960}

\bibitem[\protect\citeauthoryear{{Hill} et~al.,}{{Hill}
  et~al.}{2005}]{2005SPIE.5898..325H}
{Hill} J.~E.,  et~al., 2005, in {Siegmund} O. H.~W.,  ed.,  Society of
  Photo-Optical Instrumentation Engineers (SPIE) Conference Series Vol. 5898,
  UV, X-Ray, and Gamma-Ray Space Instrumentation for Astronomy XIV. pp
  325--340, \mn@doi{10.1117/12.618026}

\bibitem[\protect\citeauthoryear{{H{\o}g} et~al.,}{{H{\o}g}
  et~al.}{2000}]{2000A&A...355L..27H}
{H{\o}g} E.,  et~al., 2000, \aap, \href
  {http://adsabs.harvard.edu/abs/2000A%26A...355L..27H} {355, L27}

\bibitem[\protect\citeauthoryear{{Horesh} et~al.,}{{Horesh}
  et~al.}{2013}]{2013ApJ...778...63H}
{Horesh} A.,  et~al., 2013, \mn@doi [\apj] {10.1088/0004-637X/778/1/63}, \href
  {https://ui.adsabs.harvard.edu/abs/2013ApJ...778...63H} {778, 63}

\bibitem[\protect\citeauthoryear{{Hotan}, {van Straten}  \&
  {Manchester}}{{Hotan} et~al.}{2004}]{HSM04}
{Hotan} A.~W.,  {van Straten} W.,   {Manchester} R.~N.,  2004, \mn@doi [\pasa]
  {10.1071/AS04022}, \href
  {https://ui.adsabs.harvard.edu/abs/2004PASA...21..302H} {21, 302}

\bibitem[\protect\citeauthoryear{{Hyman}, {Lazio}, {Kassim}, {Ray}, {Markwardt}
   \& {Yusef-Zadeh}}{{Hyman} et~al.}{2005}]{2005Natur.434...50H}
{Hyman} S.~D.,  {Lazio} T. J.~W.,  {Kassim} N.~E.,  {Ray} P.~S.,  {Markwardt}
  C.~B.,   {Yusef-Zadeh} F.,  2005, \mn@doi [\nat] {10.1038/nature03400}, \href
  {https://ui.adsabs.harvard.edu/abs/2005Natur.434...50H} {434, 50}

\bibitem[\protect\citeauthoryear{{Hyman}, {Wijnands}, {Lazio}, {Pal},
  {Starling}, {Kassim}  \& {Ray}}{{Hyman} et~al.}{2009}]{2009ApJ...696..280H}
{Hyman} S.~D.,  {Wijnands} R.,  {Lazio} T. J.~W.,  {Pal} S.,  {Starling} R.,
  {Kassim} N.~E.,   {Ray} P.~S.,  2009, \mn@doi [\apj]
  {10.1088/0004-637X/696/1/280}, \href
  {https://ui.adsabs.harvard.edu/abs/2009ApJ...696..280H} {696, 280}

\bibitem[\protect\citeauthoryear{{Jankowski} et~al.,}{{Jankowski}
  et~al.}{2019}]{2019MNRAS.484.3691J}
{Jankowski} F.,  et~al., 2019, \mn@doi [\mnras] {10.1093/mnras/sty3390}, \href
  {https://ui.adsabs.harvard.edu/abs/2019MNRAS.484.3691J} {484, 3691}

\bibitem[\protect\citeauthoryear{{Johnston} et~al.,}{{Johnston}
  et~al.}{2008}]{2008ExA....22..151J}
{Johnston} S.,  et~al., 2008, \mn@doi [Experimental Astronomy]
  {10.1007/s10686-008-9124-7}, \href
  {http://adsabs.harvard.edu/abs/2008ExA....22..151J} {22, 151}

\bibitem[\protect\citeauthoryear{{Kniazev}, {Gvaramadze}  \&
  {Berdnikov}}{{Kniazev} et~al.}{2016}]{Kniazev2016}
{Kniazev} A.~Y.,  {Gvaramadze} V.~V.,   {Berdnikov} L.~N.,  2016, \mn@doi
  [\mnras] {10.1093/mnras/stw889}, \href
  {https://ui.adsabs.harvard.edu/abs/2016MNRAS.459.3068K} {459, 3068}

\bibitem[\protect\citeauthoryear{{Kochanek} et~al.,}{{Kochanek}
  et~al.}{2017}]{2017PASP..129j4502K}
{Kochanek} C.~S.,  et~al., 2017, \mn@doi [\pasp] {10.1088/1538-3873/aa80d9},
  \href {http://adsabs.harvard.edu/abs/2017PASP..129j4502K} {129, 104502}

\bibitem[\protect\citeauthoryear{{Kuiack}, {Huizinga}, {Molenaar}, {Prasad},
  {Rowlinson}  \& {Wijers}}{{Kuiack} et~al.}{2019}]{2019MNRAS.482.2502K}
{Kuiack} M.,  {Huizinga} F.,  {Molenaar} G.,  {Prasad} P.,  {Rowlinson} A.,
  {Wijers} R. A.~M.~J.,  2019, \mn@doi [\mnras] {10.1093/mnras/sty2810}, \href
  {https://ui.adsabs.harvard.edu/abs/2019MNRAS.482.2502K} {482, 2502}

\bibitem[\protect\citeauthoryear{{Lacy} et~al.,}{{Lacy}
  et~al.}{2019}]{2019arXiv190701981L}
{Lacy} M.,  et~al., 2019, arXiv e-prints, \href
  {https://ui.adsabs.harvard.edu/abs/2019arXiv190701981L} {p. arXiv:1907.01981}

\bibitem[\protect\citeauthoryear{{Lomb}}{{Lomb}}{1976}]{1976Ap&SS..39..447L}
{Lomb} N.~R.,  1976, \mn@doi [\apss] {10.1007/BF00648343}, \href
  {https://ui.adsabs.harvard.edu/\#abs/1976Ap&SS..39..447L} {39, 447}

\bibitem[\protect\citeauthoryear{{Maan} \& {van Leeuwen}}{{Maan} \& {van
  Leeuwen}}{2017}]{2017arXiv170906104M}
{Maan} Y.,  {van Leeuwen} J.,  2017, arXiv e-prints, \href
  {http://adsabs.harvard.edu/abs/2017arXiv170906104M} {}

\bibitem[\protect\citeauthoryear{{Marigo} et~al.,}{{Marigo}
  et~al.}{2017}]{2017ApJ...835...77M}
{Marigo} P.,  et~al., 2017, \mn@doi [\apj] {10.3847/1538-4357/835/1/77}, \href
  {http://adsabs.harvard.edu/abs/2017ApJ...835...77M} {835, 77}

\bibitem[\protect\citeauthoryear{{Marsh} et~al.,}{{Marsh}
  et~al.}{2016}]{2016Natur.537..374M}
{Marsh} T.~R.,  et~al., 2016, \mn@doi [\nat] {10.1038/nature18620}, \href
  {https://ui.adsabs.harvard.edu/abs/2016Natur.537..374M} {537, 374}

\bibitem[\protect\citeauthoryear{{Mason}, {Cropper}, {Kennedy}, {Nousek},
  {Roming}  \& {McLelland}}{{Mason} et~al.}{2000}]{2000HEAD....5.3411M}
{Mason} K.~O.,  {Cropper} M.~S.,  {Kennedy} T.~E.,  {Nousek} J.,  {Roming} P.,
   {McLelland} M.,  2000, in AAS/High Energy Astrophysics Division \#5. p.
  34.11

\bibitem[\protect\citeauthoryear{{McDonald} \& {Zijlstra}}{{McDonald} \&
  {Zijlstra}}{2015}]{2015MNRAS.448..502M}
{McDonald} I.,  {Zijlstra} A.~A.,  2015, \mn@doi [\mnras]
  {10.1093/mnras/stv007}, \href
  {https://ui.adsabs.harvard.edu/abs/2015MNRAS.448..502M} {448, 502}

\bibitem[\protect\citeauthoryear{{McDonald}, {Zijlstra}  \& {Boyer}}{{McDonald}
  et~al.}{2012}]{2012MNRAS.427..343M}
{McDonald} I.,  {Zijlstra} A.~A.,   {Boyer} M.~L.,  2012, \mn@doi [\mnras]
  {10.1111/j.1365-2966.2012.21873.x}, \href
  {http://adsabs.harvard.edu/abs/2012MNRAS.427..343M} {427, 343}

\bibitem[\protect\citeauthoryear{{McDonald}, {Zijlstra}  \&
  {Watson}}{{McDonald} et~al.}{2017}]{2017MNRAS.471..770M}
{McDonald} I.,  {Zijlstra} A.~A.,   {Watson} R.~A.,  2017, \mn@doi [\mnras]
  {10.1093/mnras/stx1433}, \href
  {http://adsabs.harvard.edu/abs/2017MNRAS.471..770M} {471, 770}

\bibitem[\protect\citeauthoryear{{McMullin}, {Waters}, {Schiebel}, {Young}  \&
  {Golap}}{{McMullin} et~al.}{2007}]{2007mcmulin}
{McMullin} J.~P.,  {Waters} B.,  {Schiebel} D.,  {Young} W.,   {Golap} K.,
  2007, in {Shaw} R.~A.,  {Hill} F.,   {Bell} D.~J.,  eds,  Astronomical
  Society of the Pacific Conference Series Vol. 376, Astronomical Data Analysis
  Software and Systems XVI. p.~127

\bibitem[\protect\citeauthoryear{{Mohan} \& {Rafferty}}{{Mohan} \&
  {Rafferty}}{2015}]{2015ascl.soft02007M}
{Mohan} N.,  {Rafferty} D.,  2015, {PyBDSF: Python Blob Detection and Source
  Finder}, Astrophysics Source Code Library (\mn@eprint {ascl} {1502.007})

\bibitem[\protect\citeauthoryear{{Mooley} et~al.,}{{Mooley}
  et~al.}{2016}]{2016ApJ...818..105M}
{Mooley} K.~P.,  et~al., 2016, \mn@doi [\apj] {10.3847/0004-637X/818/2/105},
  \href {https://ui.adsabs.harvard.edu/abs/2016ApJ...818..105M} {818, 105}

\bibitem[\protect\citeauthoryear{{Mooley} et~al.,}{{Mooley}
  et~al.}{2017}]{2017MNRAS.467L..31M}
{Mooley} K.~P.,  et~al., 2017, \mn@doi [\mnras] {10.1093/mnrasl/slw243}, \href
  {https://ui.adsabs.harvard.edu/abs/2017MNRAS.467L..31M} {467, L31}

\bibitem[\protect\citeauthoryear{{Mooley} et~al.,}{{Mooley}
  et~al.}{2018}]{2018ApJ...857..143M}
{Mooley} K.~P.,  et~al., 2018, \mn@doi [\apj] {10.3847/1538-4357/aab7f3}, \href
  {https://ui.adsabs.harvard.edu/abs/2018ApJ...857..143M} {857, 143}

\bibitem[\protect\citeauthoryear{{Morello} et~al.,}{{Morello}
  et~al.}{2019}]{2019MNRAS.483.3673M}
{Morello} V.,  et~al., 2019, \mn@doi [\mnras] {10.1093/mnras/sty3328}, \href
  {https://ui.adsabs.harvard.edu/abs/2019MNRAS.483.3673M} {483, 3673}

\bibitem[\protect\citeauthoryear{{Morris}, {Curtis}, {Douglas}, {Hawley},
  {Ag{\"u}eros}, {Bobra}  \& {Agol}}{{Morris}
  et~al.}{2018}]{2018AJ....156..203M}
{Morris} B.~M.,  {Curtis} J.~L.,  {Douglas} S.~T.,  {Hawley} S.~L.,
  {Ag{\"u}eros} M.~A.,  {Bobra} M.~G.,   {Agol} E.,  2018, \mn@doi [\aj]
  {10.3847/1538-3881/aae1ab}, \href
  {https://ui.adsabs.harvard.edu/abs/2018AJ....156..203M} {156, 203}

\bibitem[\protect\citeauthoryear{{Murphy} et~al.,}{{Murphy}
  et~al.}{2013}]{2013PASA...30....6M}
{Murphy} T.,  et~al., 2013, \mn@doi [\pasa] {10.1017/pasa.2012.006}, \href
  {https://ui.adsabs.harvard.edu/abs/2013PASA...30....6M} {30, e006}

\bibitem[\protect\citeauthoryear{{Murphy} et~al.,}{{Murphy}
  et~al.}{2017}]{2017MNRAS.466.1944M}
{Murphy} T.,  et~al., 2017, \mn@doi [\mnras] {10.1093/mnras/stw3087}, \href
  {https://ui.adsabs.harvard.edu/abs/2017MNRAS.466.1944M} {466, 1944}

\bibitem[\protect\citeauthoryear{{O'Brien}, {Rupen}, {Chomiuk}, {Ribeiro},
  {Bode}, {Sokoloski}  \& {Woudt}}{{O'Brien}
  et~al.}{2015}]{2015aska.confE..62O}
{O'Brien} T.,  {Rupen} M.,  {Chomiuk} L.,  {Ribeiro} V.,  {Bode} M.,
  {Sokoloski} J.,   {Woudt} P.~A.,  2015, in Advancing Astrophysics with the
  Square Kilometre Array (AASKA14). p.~62 (\mn@eprint {arXiv} {1502.04927})

\bibitem[\protect\citeauthoryear{{Ofek}, {Frail}, {Breslauer}, {Kulkarni},
  {Chandra}, {Gal-Yam}, {Kasliwal}  \& {Gehrels}}{{Ofek}
  et~al.}{2011}]{2011ApJ...740...65O}
{Ofek} E.~O.,  {Frail} D.~A.,  {Breslauer} B.,  {Kulkarni} S.~R.,  {Chandra}
  P.,  {Gal-Yam} A.,  {Kasliwal} M.~M.,   {Gehrels} N.,  2011, \mn@doi [\apj]
  {10.1088/0004-637X/740/2/65}, \href
  {https://ui.adsabs.harvard.edu/abs/2011ApJ...740...65O} {740, 65}

\bibitem[\protect\citeauthoryear{{Offringa}}{{Offringa}}{2010}]{2010offringa}
{Offringa} A.~R.,  2010, {AOFlagger: RFI Software} (\mn@eprint {ascl}
  {1010.017})

\bibitem[\protect\citeauthoryear{{Osten}}{{Osten}}{2008}]{2008arXiv0801.2573O}
{Osten} R.~A.,  2008, arXiv e-prints, \href
  {https://ui.adsabs.harvard.edu/abs/2008arXiv0801.2573O} {p. arXiv:0801.2573}

\bibitem[\protect\citeauthoryear{{Osten} \& {Bastian}}{{Osten} \&
  {Bastian}}{2008}]{2008ApJ...674.1078O}
{Osten} R.~A.,  {Bastian} T.~S.,  2008, \mn@doi [\apj] {10.1086/525013}, \href
  {https://ui.adsabs.harvard.edu/abs/2008ApJ...674.1078O} {674, 1078}

\bibitem[\protect\citeauthoryear{{Pepper} et~al.,}{{Pepper}
  et~al.}{2007}]{2007PASP..119..923P}
{Pepper} J.,  et~al., 2007, \mn@doi [Publications of the Astronomical Society
  of the Pacific] {10.1086/521836}, \href
  {https://ui.adsabs.harvard.edu/abs/2007PASP..119..923P} {119, 923}

\bibitem[\protect\citeauthoryear{{Perera} et~al.,}{{Perera}
  et~al.}{2018}]{2018MNRAS.478..218P}
{Perera} B.~B.~P.,  et~al., 2018, \mn@doi [\mnras] {10.1093/mnras/sty1116},
  \href {http://adsabs.harvard.edu/abs/2018MNRAS.478..218P} {478, 218}

\bibitem[\protect\citeauthoryear{{Pickles} \& {Depagne}}{{Pickles} \&
  {Depagne}}{2010}]{2010PASP..122.1437P}
{Pickles} A.,  {Depagne} {\'E}.,  2010, \mn@doi [\pasp] {10.1086/657947}, \href
  {https://ui.adsabs.harvard.edu/abs/2010PASP..122.1437P} {122, 1437}

\bibitem[\protect\citeauthoryear{{Pietka}, {Fender}  \& {Keane}}{{Pietka}
  et~al.}{2015}]{2015MNRAS.446.3687P}
{Pietka} M.,  {Fender} R.~P.,   {Keane} E.~F.,  2015, \mn@doi [\mnras]
  {10.1093/mnras/stu2335}, \href
  {https://ui.adsabs.harvard.edu/abs/2015MNRAS.446.3687P} {446, 3687}

\bibitem[\protect\citeauthoryear{{Pojmanski}}{{Pojmanski}}{1997}]{1997AcA....47..467P}
{Pojmanski} G.,  1997, \actaa, \href
  {http://adsabs.harvard.edu/abs/1997AcA....47..467P} {47, 467}

\bibitem[\protect\citeauthoryear{{Pojmanski}}{{Pojmanski}}{2002}]{2002AcA....52..397P}
{Pojmanski} G.,  2002, \actaa, \href
  {http://adsabs.harvard.edu/abs/2002AcA....52..397P} {52, 397}

\bibitem[\protect\citeauthoryear{{Polisensky} et~al.,}{{Polisensky}
  et~al.}{2016}]{2016ApJ...832...60P}
{Polisensky} E.,  et~al., 2016, \mn@doi [\apj] {10.3847/0004-637X/832/1/60},
  \href {https://ui.adsabs.harvard.edu/abs/2016ApJ...832...60P} {832, 60}

\bibitem[\protect\citeauthoryear{{Potter} et~al.,}{{Potter}
  et~al.}{2010}]{Potter2010}
{Potter} S.~B.,  et~al., 2010, \mn@doi [\mnras]
  {10.1111/j.1365-2966.2009.15944.x}, \href
  {https://ui.adsabs.harvard.edu/abs/2010MNRAS.402.1161P} {402, 1161}

\bibitem[\protect\citeauthoryear{{Prasad} et~al.,}{{Prasad}
  et~al.}{2016}]{2016JAI.....541008P}
{Prasad} P.,  et~al., 2016, \mn@doi [Journal of Astronomical Instrumentation]
  {10.1142/S2251171716410087}, \href
  {https://ui.adsabs.harvard.edu/abs/2016JAI.....541008P} {5, 1641008}

\bibitem[\protect\citeauthoryear{{Price-Whelan} et~al.,}{{Price-Whelan}
  et~al.}{2018}]{astropy:2018}
{Price-Whelan} A.~M.,  et~al., 2018, \mn@doi [\aj] {10.3847/1538-3881/aabc4f},
  \href {https://ui.adsabs.harvard.edu/#abs/2018AJ....156..123T} {156, 123}

\bibitem[\protect\citeauthoryear{{Rajwade}, {Chennamangalam}, {Lorimer}  \&
  {Karastergiou}}{{Rajwade} et~al.}{2018}]{RCJ+18}
{Rajwade} K.,  {Chennamangalam} J.,  {Lorimer} D.,   {Karastergiou} A.,  2018,
  \mn@doi [\mnras] {10.1093/mnras/sty1695}, \href
  {https://ui.adsabs.harvard.edu/abs/2018MNRAS.479.3094R} {479, 3094}

\bibitem[\protect\citeauthoryear{{Ransom}, {Eikenberry}  \&
  {Middleditch}}{{Ransom} et~al.}{2002}]{REM02}
{Ransom} S.~M.,  {Eikenberry} S.~S.,   {Middleditch} J.,  2002, \mn@doi [\aj]
  {10.1086/342285}, \href {http://adsabs.harvard.edu/abs/2002AJ....124.1788R}
  {124, 1788}

\bibitem[\protect\citeauthoryear{{Richards}, {Starr}, {Miller}, {Bloom},
  {Butler}, {Brink}  \& {Crellin-Quick}}{{Richards}
  et~al.}{2012}]{2012ApJS..203...32R}
{Richards} J.~W.,  {Starr} D.~L.,  {Miller} A.~A.,  {Bloom} J.~S.,  {Butler}
  N.~R.,  {Brink} H.,   {Crellin-Quick} A.,  2012, \mn@doi [The Astrophysical
  Journal Supplement Series] {10.1088/0067-0049/203/2/32}, \href
  {https://ui.adsabs.harvard.edu/\#abs/2012ApJS..203...32R} {203, 32}

\bibitem[\protect\citeauthoryear{{Robitaille} \& {Bressert}}{{Robitaille} \&
  {Bressert}}{2012}]{2012ascl.soft08017R}
{Robitaille} T.,  {Bressert} E.,  2012, {APLpy: Astronomical Plotting Library
  in Python}, Astrophysics Source Code Library (\mn@eprint {ascl} {1208.017})

\bibitem[\protect\citeauthoryear{{Roy}, {Hyman}, {Pal}, {Lazio}, {Ray}  \&
  {Kassim}}{{Roy} et~al.}{2010}]{2010ApJ...712L...5R}
{Roy} S.,  {Hyman} S.~D.,  {Pal} S.,  {Lazio} T. J.~W.,  {Ray} P.~S.,
  {Kassim} N.~E.,  2010, \mn@doi [\apjl] {10.1088/2041-8205/712/1/L5}, \href
  {https://ui.adsabs.harvard.edu/abs/2010ApJ...712L...5R} {712, L5}

\bibitem[\protect\citeauthoryear{{Scargle}}{{Scargle}}{1982}]{1982ApJ...263..835S}
{Scargle} J.~D.,  1982, \mn@doi [\apj] {10.1086/160554}, \href
  {https://ui.adsabs.harvard.edu/\#abs/1982ApJ...263..835S} {263, 835}

\bibitem[\protect\citeauthoryear{{Schinckel}, {Bunton}, {Cornwell}, {Feain}  \&
  {Hay}}{{Schinckel} et~al.}{2012}]{2012SPIE.8444E..2AS}
{Schinckel} A.~E.,  {Bunton} J.~D.,  {Cornwell} T.~J.,  {Feain} I.,   {Hay}
  S.~G.,  2012, in Ground-based and Airborne Telescopes IV. p. 84442A,
  \mn@doi{10.1117/12.926959}

\bibitem[\protect\citeauthoryear{{Seaquist}}{{Seaquist}}{1977}]{1977ApJ...211..547S}
{Seaquist} E.~R.,  1977, \mn@doi [\apj] {10.1086/154961}, \href
  {https://ui.adsabs.harvard.edu/abs/1977ApJ...211..547S} {211, 547}

\bibitem[\protect\citeauthoryear{{Shappee} et~al.,}{{Shappee}
  et~al.}{2014}]{2014ApJ...788...48S}
{Shappee} B.~J.,  et~al., 2014, \mn@doi [\apj] {10.1088/0004-637X/788/1/48},
  \href {http://adsabs.harvard.edu/abs/2014ApJ...788...48S} {788, 48}

\bibitem[\protect\citeauthoryear{{Smirnov} \& {Tasse}}{{Smirnov} \&
  {Tasse}}{2015}]{2015tasse}
{Smirnov} O.~M.,  {Tasse} C.,  2015, \mn@doi [\mnras] {10.1093/mnras/stv418},
  \href {https://ui.adsabs.harvard.edu/#abs/2015MNRAS.449.2668S} {449, 2668}

\bibitem[\protect\citeauthoryear{{Sood} \& {Campbell-Wilson}}{{Sood} \&
  {Campbell-Wilson}}{1994}]{1994IAUC.6006....1S}
{Sood} R.,  {Campbell-Wilson} D.,  1994, International Astronomical Union
  Circular, \href {https://ui.adsabs.harvard.edu/abs/1994IAUC.6006....1S}
  {6006, 1}

\bibitem[\protect\citeauthoryear{{Stewart} et~al.,}{{Stewart}
  et~al.}{2016}]{2016MNRAS.456.2321S}
{Stewart} A.~J.,  et~al., 2016, \mn@doi [\mnras] {10.1093/mnras/stv2797}, \href
  {https://ui.adsabs.harvard.edu/abs/2016MNRAS.456.2321S} {456, 2321}

\bibitem[\protect\citeauthoryear{{Swinbank} et~al.,}{{Swinbank}
  et~al.}{2015}]{Swinbank2015}
{Swinbank} J.~D.,  et~al., 2015, \mn@doi [Astronomy and Computing]
  {10.1016/j.ascom.2015.03.002}, \href
  {http://adsabs.harvard.edu/abs/2015A%26C....11...25S} {11, 25}

\bibitem[\protect\citeauthoryear{{Takata}, {Hu}, {Lin}, {Tam}, {Pal}, {Hui},
  {Kong}  \& {Cheng}}{{Takata} et~al.}{2018}]{2018ApJ...853..106T}
{Takata} J.,  {Hu} C.~P.,  {Lin} L.~C.~C.,  {Tam} P.~H.~T.,  {Pal} P.~S.,
  {Hui} C.~Y.,  {Kong} A.~K.~H.,   {Cheng} K.~S.,  2018, \mn@doi [\apj]
  {10.3847/1538-4357/aaa23d}, \href
  {https://ui.adsabs.harvard.edu/abs/2018ApJ...853..106T} {853, 106}

\bibitem[\protect\citeauthoryear{{Tang}, {Bressan}, {Rosenfield}, {Slemer},
  {Marigo}, {Girardi}  \& {Bianchi}}{{Tang} et~al.}{2014}]{2014MNRAS.445.4287T}
{Tang} J.,  {Bressan} A.,  {Rosenfield} P.,  {Slemer} A.,  {Marigo} P.,
  {Girardi} L.,   {Bianchi} L.,  2014, \mn@doi [\mnras]
  {10.1093/mnras/stu2029}, \href
  {https://ui.adsabs.harvard.edu/abs/2014MNRAS.445.4287T} {445, 4287}

\bibitem[\protect\citeauthoryear{{Tasse} et~al.,}{{Tasse}
  et~al.}{2018}]{2018tasse}
{Tasse} C.,  et~al., 2018, \mn@doi [\aap] {10.1051/0004-6361/201731474}, \href
  {https://ui.adsabs.harvard.edu/#abs/2018A&A...611A..87T} {611}

\bibitem[\protect\citeauthoryear{{Thyagarajan}, {Helfand}, {White}  \&
  {Becker}}{{Thyagarajan} et~al.}{2011}]{2011ApJ...742...49T}
{Thyagarajan} N.,  {Helfand} D.~J.,  {White} R.~L.,   {Becker} R.~H.,  2011,
  \mn@doi [\apj] {10.1088/0004-637X/742/1/49}, \href
  {https://ui.adsabs.harvard.edu/abs/2011ApJ...742...49T} {742, 49}

\bibitem[\protect\citeauthoryear{{Tingay} et~al.,}{{Tingay}
  et~al.}{2012}]{2012rsri.confE..36T}
{Tingay} S.,  et~al., 2012, in Resolving The Sky - Radio Interferometry: Past,
  Present and Future. p.~36 (\mn@eprint {arXiv} {1212.1327})

\bibitem[\protect\citeauthoryear{{TraP contributors}}{{TraP
  contributors}}{2014}]{TraP}
{TraP contributors} 2014, {TraP: Transients discovery pipeline for image-plane
  surveys}, Astrophysics Source Code Library (\mn@eprint {ascl} {1412.011})

\bibitem[\protect\citeauthoryear{{Trigilio}, {Umana}  \& {Migenes}}{{Trigilio}
  et~al.}{1993}]{1993MNRAS.260..903T}
{Trigilio} C.,  {Umana} G.,   {Migenes} V.,  1993, \mn@doi [\mnras]
  {10.1093/mnras/260.4.903}, \href
  {https://ui.adsabs.harvard.edu/abs/1993MNRAS.260..903T} {260, 903}

\bibitem[\protect\citeauthoryear{{Varghese}, {Obenberger}, {Dowell}  \&
  {Taylor}}{{Varghese} et~al.}{2019}]{2019ApJ...874..151V}
{Varghese} S.~S.,  {Obenberger} K.~S.,  {Dowell} J.,   {Taylor} G.~B.,  2019,
  \mn@doi [\apj] {10.3847/1538-4357/ab07c6}, \href
  {https://ui.adsabs.harvard.edu/abs/2019ApJ...874..151V} {874, 151}

\bibitem[\protect\citeauthoryear{{Villadsen} \& {Hallinan}}{{Villadsen} \&
  {Hallinan}}{2019}]{2019ApJ...871..214V}
{Villadsen} J.,  {Hallinan} G.,  2019, \mn@doi [\apj]
  {10.3847/1538-4357/aaf88e}, \href
  {https://ui.adsabs.harvard.edu/abs/2019ApJ...871..214V} {871, 214}

\bibitem[\protect\citeauthoryear{{Walter}, {Neff}, {Gibson}, {Linsky},
  {Rodono}, {Gary}  \& {Butler}}{{Walter} et~al.}{1987}]{1987A&A...186..241W}
{Walter} F.~M.,  {Neff} J.~E.,  {Gibson} D.~M.,  {Linsky} J.~L.,  {Rodono} M.,
  {Gary} D.~E.,   {Butler} C.~J.,  1987, \aap, \href
  {https://ui.adsabs.harvard.edu/abs/1987A&A...186..241W} {186, 241}

\bibitem[\protect\citeauthoryear{{Wang}, {Manchester}  \& {Johnston}}{{Wang}
  et~al.}{2007}]{2007MNRAS.377.1383W}
{Wang} N.,  {Manchester} R.~N.,   {Johnston} S.,  2007, \mn@doi [\mnras]
  {10.1111/j.1365-2966.2007.11703.x}, \href
  {https://ui.adsabs.harvard.edu/abs/2007MNRAS.377.1383W} {377, 1383}

\bibitem[\protect\citeauthoryear{{Wray}, {Eyer}  \& {Paczy{\'n}ski}}{{Wray}
  et~al.}{2004}]{2004MNRAS.349.1059W}
{Wray} J.~J.,  {Eyer} L.,   {Paczy{\'n}ski} B.,  2004, \mn@doi [\mnras]
  {10.1111/j.1365-2966.2004.07587.x}, \href
  {https://ui.adsabs.harvard.edu/\#abs/2004MNRAS.349.1059W} {349, 1059}

\bibitem[\protect\citeauthoryear{{Zauderer} et~al.,}{{Zauderer}
  et~al.}{2011}]{2011Natur.476..425Z}
{Zauderer} B.~A.,  et~al., 2011, \mn@doi [\nat] {10.1038/nature10366}, \href
  {https://ui.adsabs.harvard.edu/abs/2011Natur.476..425Z} {476, 425}

\bibitem[\protect\citeauthoryear{{Zic} et~al.,}{{Zic}
  et~al.}{2019}]{2019arXiv190606570Z}
{Zic} A.,  et~al., 2019, arXiv e-prints, \href
  {https://ui.adsabs.harvard.edu/abs/2019arXiv190606570Z} {p. arXiv:1906.06570}

\bibitem[\protect\citeauthoryear{{van Haarlem} et~al.,}{{van Haarlem}
  et~al.}{2013}]{2013A&A...556A...2V}
{van Haarlem} M.~P.,  et~al., 2013, \mn@doi [\aap]
  {10.1051/0004-6361/201220873}, \href
  {https://ui.adsabs.harvard.edu/abs/2013A&A...556A...2V} {556, A2}

\makeatother
\end{thebibliography}



 \appendix

 \section{MKT J170456.2$-$482100 MeerKAT flux density measurements}
 \label{App: flux measurements}
 
 \begin{table}
     \centering
     \begin{tabular}{p{0.9cm}p{1.4cm}p{1.1cm}p{1.7cm}p{1.4cm}}
     \hline \\
        Epoch (MJD) & Epoch (date) & Flux density (mJy) & Uncertainty (mJy) & Local RMS noise (mJy) \\
        \hline \\
        58068 & 2017-11-11 & 0.048* & 0.026 & 0.036 \\
        58222 & 2018-04-14 & 0.156 & 0.018 & 0.018 \\
        58369 & 2018-09-08 & 0.043* & 0.026 & 0.039 \\
        58375 & 2018-09-14 & 0.189 & 0.025 & 0.026 \\
        58382 & 2018-09-21 & 0.116 & 0.024 & 0.026 \\
        58389 & 2018-09-28 & 0.339 & 0.034 & 0.035 \\
        58396 & 2018-10-05 & 0.137 & 0.027 & 0.028 \\
        58402 & 2018-10-11 & 0.227 & 0.026 & 0.027 \\
        58403 & 2018-10-12 & 0.355 & 0.043 & 0.044 \\
        58410 & 2018-10-19 & 0.152 & 0.028 & 0.028 \\
        58418 & 2018-10-27 & 0.098 & 0.025 & 0.026 \\
        58425 & 2018-11-03 & 0.113 & 0.026 & 0.027 \\
        58432 & 2018-11-10 & 0.140 & 0.026 & 0.027 \\
        58439 & 2018-11-17 & -0.027* & 0.040 & 0.031 \\
        58446 & 2018-11-24 & 0.067 & 0.027 & 0.033 \\
        58454 & 2018-12-02 & 0.096 & 0.030 & 0.031 \\
        58460 & 2018-12-08 & 0.136 & 0.029 & 0.031 \\
        58467 & 2018-12-15 & 0.198 & 0.031 & 0.031 \\
        58474 & 2018-12-22 & 0.202 & 0.036 & 0.036 \\
        58481 & 2018-12-29 & 0.189 & 0.035 & 0.036 \\
        58488 & 2019-01-05 & 0.192 & 0.030 & 0.030 \\
        58495 & 2019-01-12 & 0.164 & 0.031 & 0.032 \\
        58502 & 2019-01-19 & -0.005* & 0.006 & 0.028 \\
        58509 & 2019-01-26 & 0.211 & 0.029 & 0.029 \\
        58515 & 2019-02-01 & 0.025* & 0.017 & 0.030 \\
        58523 & 2019-02-09 & 0.119 & 0.033 & 0.035 \\
        58530 & 2019-02-16 & 0.077 & 0.026 & 0.029 \\
        58537 & 2019-02-23 & 0.115 & 0.029 & 0.031 \\
        58543 & 2019-03-01 & 0.103 & 0.027 & 0.028 \\
        58551 & 2019-03-09 & 0.091 & 0.028 & 0.031 \\
        58560 & 2019-03-18 & 0.038* & 0.023 & 0.035 \\
        58567 & 2019-03-25 & -0.112* & 0.408 & 0.032 \\
        58574 & 2019-04-01 & -0.116* & 0.373 & 0.037 \\
        58582 & 2019-04-09 & 0.108 & 0.036 & 0.041 \\
        58588 & 2019-04-15 & 0.084 & 0.028 & 0.031 \\
        58593 & 2019-04-20 & 0.032* & 0.020 & 0.032 \\
        58602 & 2019-04-29 & 0.059* & 0.029 & 0.038 \\
        58607 & 2019-05-04 & 0.033* & 0.020 & 0.031 \\
        58614 & 2019-05-11 & 0.109 & 0.032 & 0.034 \\
        58621 & 2019-05-18 & 0.097 & 0.035 & 0.039 \\
        58628 & 2019-05-25 & 0.105 & 0.031 & 0.033 \\
        58634 & 2019-05-31 & 0.113 & 0.034 & 0.036 \\
        58642 & 2019-06-08 & 0.015* & 0.012 & 0.040 \\
        58650 & 2019-06-16 & 0.115 & 0.032 & 0.035 \\
        58658 & 2019-06-24 & 0.051* & 0.024 & 0.031 \\
        58664 & 2019-06-30 & 2.122* & 150.234 & 0.031 \\
        58671 & 2019-07-07 & -0.002* & 0.002 & 0.030 \\
        58678 & 2019-07-14 & 0.179 & 0.030 & 0.030 \\
        \hline \\
     \end{tabular}
     \caption{Flux densities measured by MeerKAT for each epoch for \fb. These are the original, unscaled (the primary beam correction of $2.00\pm0.02$ has not been applied) peak flux density measurements from \trap, (*) indicates measurements that were plotted as upper limits in Figures\,\ref{fig: Radio variability} and \ref{fig: recent radio optical}.}
     \label{tab: radio flux measurements}
 \end{table}


\bsp	
\label{lastpage}
\end{document}